\documentclass[aps,prl,amsmath,twocolumn,groupedaddress,letterpaper,floatfix]{revtex4-1}
\usepackage{graphicx,color}
\usepackage{verbatim}
\usepackage{amssymb}   
\usepackage{amsmath}
\usepackage{amsfonts}
\usepackage{mathdots}
\usepackage{hyperref}
\usepackage{epsfig}
\usepackage{braket}
\usepackage{bm}
 
 \newcommand{\addYM}[1]{\textcolor{black}{ #1}}

\begin{document}
	\title{Josephson Diode Effect Induced by Valley Polarization in Twisted Bilayer Graphene}
	\author{Jin-Xin Hu}\thanks{These authors contributed equally to this work}
	\author{Zi-Ting Sun}\thanks{These authors contributed equally to this work}
    \author{Ying-Ming Xie} \thanks{Corresponding author: ymxi@ust.hk}
	\author{K. T. Law} \thanks{Corresponding author: phlaw@ust.hk}
	\affiliation{Department of Physics, Hong Kong University of Science and Technology, Clear Water Bay, Hong Kong, China} 		
	\date{\today}
	\begin{abstract}
		Recently, the Josephson diode effect (JDE), in which the superconducting critical current magnitudes differ when the currents flow in opposite directions, has attracted great interest. In particular, it was demonstrated that gate-defined Josephson junctions based on magic-angle twisted bilayer graphene showed a strong nonreciprocal effect when the weak-link region is gated to a correlated insulating state at half-filling (two holes per moir\'e cell). However, the mechanism behind such a phenomenon is not yet understood. In this work, we show that the interaction-driven valley polarization, together with the trigonal warping of the Fermi surface, induce the JDE. The valley polarization, which lifts the degeneracy of the states in the two valleys, induces a relative phase difference between the first and the second harmonics of supercurrent and results in the JDE. We further show that the nontrivial current phase relation, which is responsible for the JDE, also generates the asymmetric Shapiro steps.
	\end{abstract}
	\pacs{}	
	\maketitle

\emph{Introduction.}---Supercurrents flow through a junction formed by two superconductors connected by a weak link, which are called Josephson junctions (JJs) \cite{josephson1962possible,josephson1964coupled,anderson1963probable,ambegaokar1963tunneling}. Symmetry breaking plays a key role in the properties of JJs. For example, $\pi$-JJs can be formed when the time-reversal symmetry is broken, which exhibit a phase difference of $\pi$ for the two superconductors in the ground state \cite{feofanov2010implementation,yamashita2005superconducting,kato2007decoherence,yamashita2006superconducting}. When both time-reversal and inversion symmetry are broken, JJs can show the Josephson diode effect (JDE) \cite{dolcini2015topological,chen2018asymmetric,davydova2022universal,zhang2022general,tanaka2022theory,lu2022josephson,wang2022symmetry}, in which the critical supercurrent $|I_{c}|$ is nonreciprocal in the sense that $|I_{c+}|$ for the current flowing in the `$+$' direction is different from $|I_{c-}|$ for the opposite `$-$' direction. Such nonreciprocity in supercurrents could have potential applications in superconducting electronics \cite{ando2020observation,misaki2021theory,rymarz2021hardware}. Recently, there has been worldwide interest in exploring the JDE in various systems, such as in NbSe$_2$/Nb$_3$Br$_8$/NbSe$_2$ heterostructures \cite{wu2022field}, topological semimetals \cite{pal2022josephson} and gated-defined JJs in twisted bilayer graphene (TBG) \cite{diez2021magnetic}.

The observation of JDE in gated-defined JJs based on TBG is particularly interesting \cite{diez2021magnetic}. In the experiment, a single piece of magic-angle TBG was gated into three different regions to form a superconductor/correlated insulator/superconductor JJ as depicted in Fig.\ref{fig:fig1}~(a). When the correlated insulating state is at half-filling (two holes per moir\'e unit cell), a large JDE was observed. It is important to note that current theories of JDE \cite{dolcini2015topological,chen2018asymmetric,tanaka2022theory,lu2022josephson,zhang2022general} require the presence of spin-orbit coupling, but the spin-orbit coupling is negligible in TBG. Moreover, external in-plane magnetic fields were required to induce JDE in other recent experiments \cite{pal2022josephson,baumgartner2022supercurrent}. It was proposed that the in-plane magnetic field induces finite-momentum Cooper pairing at the surface of the superconductors, which is essential for explaining the JDE \cite{davydova2022universal}. On the other hand, in gate-defined JJ in TBG, time-reversal symmetry is broken spontaneously at the weak-link region by interactions and there is no evidence of finite-momentum pairing in the superconducting regions. Therefore, a new microscopic theory is needed to understand this interaction-driven JDE in TBG.

\begin{figure}
		\centering
		\includegraphics[width=1.0\linewidth]{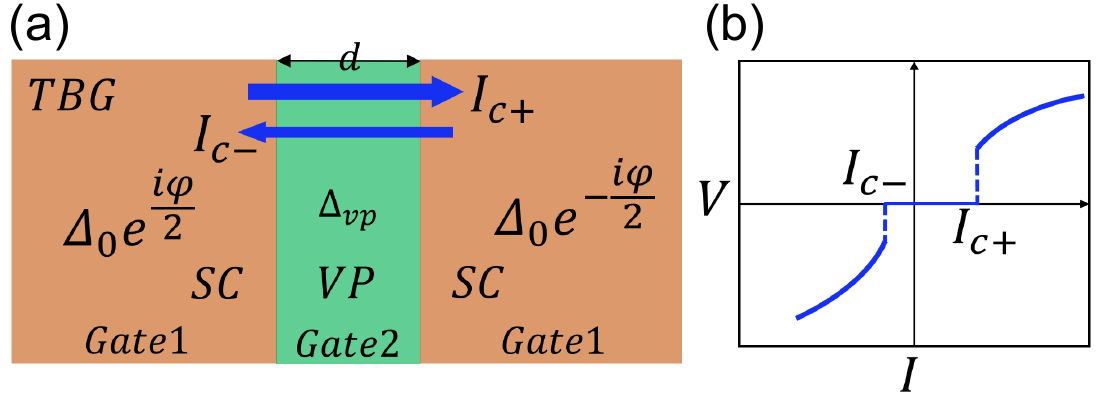}
		\caption{(a) A schematic picture of a gate-defined JJ based on magic-angle TBG. The left (right) side of the junction is superconducting (SC) with pairing order parameter $\Delta_0e^{\pm i\varphi/2}$, respectively. The weak-link region is an interaction-driven valley polarized (VP) state with width $d$. (b) A schematic illustration of the $V-I$ curve of a JJ with asymmetric critical currents $|I_{c+}|\neq |I_{c-}|$, where $V$ is the voltage across the JJ.}
		\label{fig:fig1}
\end{figure}

In this work, we show that the interaction-driven valley polarization order parameter at the weak-link region [Fig.\ref{fig:fig1}~(a)], as well as the trigonal warping of the Fermi surface [Fig.\ref{fig:fig3}~(b)] play essential roles in inducing the JDE. In the following sections, we first introduce a continuum model describing the gate-defined JJ with a valley-polarized state as the weak link. In the one-dimensional (1D) limit, we show analytically how the valley polarization, together with the trigonal warping of the Fermi surface, induce a relative phase difference between the first and the second harmonics of the Josephson current as shown in Eq. (\ref{Eq6}). This nontrivial current-phase relation (CPR) gives rise to JDE. Second, we illustrate the JDE for magic-angle TBG numerically with a lattice model. Third, we show that gate-defined JJs would also exhibit asymmetric Shapiro steps (Fig.\ref{fig:fig4}) which share the same origin as the JDE. Importantly, our theory can be generalized to JJs with magnetic field-driven spin polarization and it provides an alternative explanation of JDEs observed in other recent experiments \cite{pal2022josephson,baumgartner2022supercurrent,jeon2022zero}.

\emph{Continuum Model.}---For magic-angle TBG with valley degrees of freedom with the interaction-induced valley polarization and trigonally warped Fermi surfaces, the low-energy effective Hamiltonian has the form \cite{yuan2018model,xie2022valley}
\begin{equation}
\label{Eq1}
H_{\tau}^{2D}=\lambda_0 (k_x^2+k_y^2 )+\tau \lambda_1 k_x (k_x^2-3 k_y^2 )+\tau \Delta_{vp}-\mu,
\end{equation}
where $\tau=\pm1$ is the valley index. The $\lambda_0$ term is the kinetic energy, while the $\lambda_1$ term denotes the trigonal warping effect, which breaks intra-valley inversion symmetry such that $H_{\tau}^{2D}(k_x)\neq H_{\tau}^{2D}(-k_x)$. The time-reversal symmetry is also broken by the valley polarization $\Delta_{vp}$. For simplicity, we first take $k_y=0$ such that
\begin{equation}
\label{Eq2}
h_\tau=\lambda_0 k_x^2+\tau \lambda_1 k_x^3+\tau \Delta_{vp}-\mu.
\end{equation}
This effective 1D model allows the key results to be calculated analytically and the properties of the two-dimensional system will be demonstrated using a lattice model numerically in a later section.

In Fig.\ref{fig:fig2}~(a), a schematic figure of a 1D superconductor/valley polarized state/superconductor (SC-VP-SC) JJ is shown, where $\varphi$ is the phase difference between the two superconductors. We assume that the pairing in the superconducting regions are conventional $s$-wave pairing, and the weak-link region is a valley-polarized state ($\Delta_{vp}$ is finite which breaks the degeneracy of the two valleys).  The energy bands of $h_\tau$ are shown in Fig.\ref{fig:fig2}~(b).  As Andreev reflections only involve electrons near the Fermi surface, we can make the Andreev approximation to linearize the dispersion relations in the vicinity of the Fermi momentum, and the full junction can be described by the BDG Hamiltonian in the Nambu basis as
$[\psi_{\tau\alpha}(x),\psi^\dagger_{-\tau,-\alpha}(x)]^T$
\begin{equation}
\label{Eq3}
\hat{H}_{\alpha}^{\tau}=\left(
\begin{matrix}{}
  &\hat{h}_{\tau,\alpha}(x)  & \Delta_s(x)   \\
  &\Delta_s^{*}(x)  &  -\hat{h}_{-\tau,-\alpha}^{*}(x)
\end{matrix}\right),
\end{equation}
where $\hat{h}_{\tau,\alpha}(x)=-i\hbar v_{\tau,\alpha}(x)\partial_x+\tau\Delta_{vp}(x)$, and the Fermi velocity along the current direction is given by $v_{\tau,\alpha}(x)=v_{s,\tau\alpha}[\Theta(-x)+\Theta(x-d)]+v_{vp,\tau\alpha}\Theta(x)\Theta(d-x)$, where $v_{s, \tau \alpha}$ and $v_{vp, \tau \alpha}$ are the Fermi velocities for the superconducting and the valley-polarized regions, respectively. Here, $\alpha=\pm1$ denote the right and left movers of the electrons. The slopes of the black arrows in Fig.\ref{fig:fig2}~(b) indicate the Fermi velocities of the left and right movers schematically. Notably, the trigonal warping term which breaks the intra-valley inversion symmetry leads to $v_{vp, \tau +} \neq -v_{vp, \tau-}$. To describe the SC-VP-SC junction, the superconducting order parameter is set to be $\Delta_s(x)=\Delta_0[e^{i\varphi/2}\Theta(-x)+e^{-i\varphi/2}\Theta(x-d)]$, and the valley polarization order parameter is $\Delta_{vp}(x)=\Delta_{vp}\Theta(x)\Theta(d-x)$. In the calculations for Fig.\ref{fig:fig2}, model parameters are $\lambda_0=0.5$ eV$\cdot$nm$^2$, $\lambda_1=0.2$ eV$\cdot$nm$^3$, $\mu=0.15$ eV, $\Delta_0=4$ meV, $d=40$ nm.

\begin{figure}
		\centering
		\includegraphics[width=1.0\linewidth]{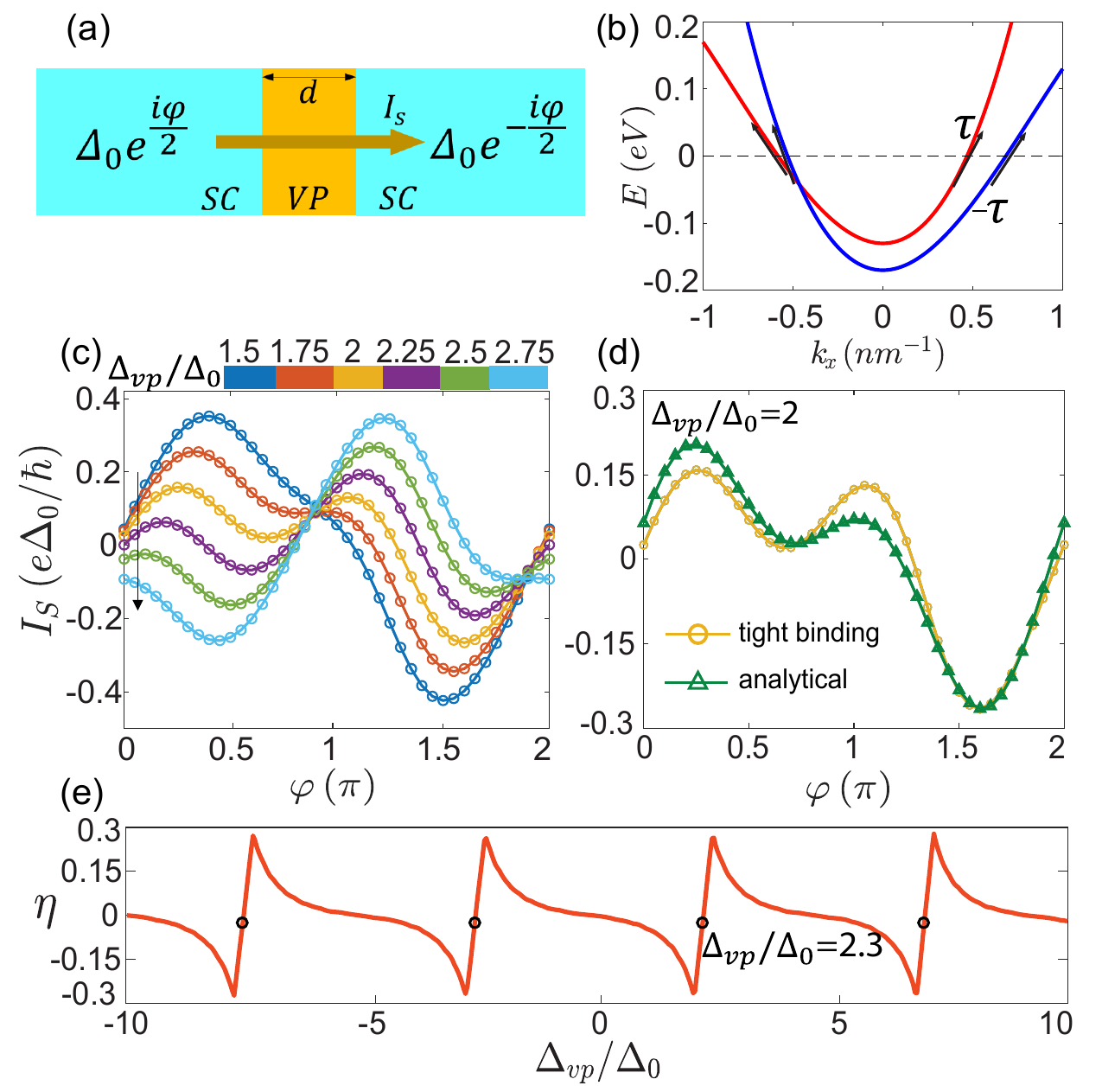}
		\caption{(a) Schematic illustration of a 1D JJ. The left (right) side of the junction is a conventional superconductor with order parameter $\Delta_0e^{+(-)i\varphi/2}$. The weak-link region is valley polarized with width $d$. (b) The band structure of the weak-link region with valley polarization order parameter $\Delta_{vp}=20$ meV. The slopes of the black arrows indicate the amplitudes of the Fermi velocities at the Fermi energy. (c) The Josephson CPR $I_s$ for $\Delta_{vp}/\Delta_0$ going from $1.5$ to $2.75$. As $\Delta_{vp}/\Delta_0$ increases (indicated by the black arrow), $\partial I_s/\partial \varphi$ at $\varphi=0^+$ changes sign. (d) The tight binding (yellow circle) and analytical (green triangle) calculations of $I_s$ for $\Delta_{vp}/\Delta_0=2$. (e) The nonreciprocity efficiency $\eta$ as a function of $\Delta_{vp}/\Delta_0$. The 0-$\pi$ transition points are labelled by black circles. The temperature is set to be $k_B T=0.2\Delta_0$.}
		\label{fig:fig2}
\end{figure}
\emph{Nonreciprocal CPR.}---A lattice model of Eq. (\ref{Eq2}) is established in the Supplementary Material \cite{NoteX}, and the Josephson supercurrent $I_s$ passing through the JJ with different $\Delta_{vp}$ are calculated \cite{NoteX} and shown in Fig.\ref{fig:fig2}~(c). It is interesting to note that as $\Delta_{vp}/\Delta_0$ increases (say, from 1.5 to 2.75), $\partial I_s/\partial \varphi$ at $\varphi=0^+$ changes sign. As a result, the CPR of $I_s$ as a function of $\varphi$ changes from $I_s \approx \sin \varphi$ to $I_s \approx \sin (\varphi + \pi) $. In another word, there is a 0 to $\pi$-junction transition as $\Delta_{vp}$ increases. Importantly, near the $0$-$\pi$ transition, the critical (or the maximum) supercurrent flowing in the positive direction $I_{c+}$ differs from the critical supercurrent $I_{c-}$ flowing in the negative direction. The nonreciprocity efficiency $\eta = (I_{c+} - |I_{c-}|)/(I_{c+} + |I_{c-}|)$ can be as large as $30\%$ at $\Delta_{vp}/\Delta_0=2$ [yellow circle line in Fig.\ref{fig:fig2}~(c)]. In other words, the JJ with valley polarization and the trigonal warping term shows a significant JDE near the 0-$\pi$ transition.

In the remainder of this section, we present an analytical description of JDE. To calculate the Josephson current, we use \cite{brouwer1997anomalous}
\begin{equation}
\label{Eq4}
I_s(\varphi)=-\frac{4e}{\hbar\beta}\frac{d}{d\varphi}\sum_{n=0}^{\infty}\ln \text{det}[1-S_A(i\omega_n,\varphi)S_N(i\omega_n,\varphi)],
\end{equation}
where $\beta=1/k_BT$, $T$ is the temperature and the Matsubara frequencies $\omega_n=(2n+1)\pi/\beta$. $S_A$, $S_N$ are the scattering matrices of the junction for the Andreev reflection and the normal scattering processes, respectively. As shown in the Supplementary Material \cite{NoteX}, we find that the 0-$\pi$ transitions occur at
\begin{equation}
\label{Eq5}
\frac{2\Delta_{vp}}{E_T}=(m+\frac{1}{2})\pi,\  m \in \mathbb{Z}.
\end{equation}
Here, $E_T$ is the Thouless energy measuring the bandwidth, which is defined as $E_T=\hbar \bar{v}_{vp}/d$, and $\bar{v}_{vp}=2/( v_{vp,++}^{-1}+ v_{vp,-+}^{-1})$. Note that $v_{vp, \tau \alpha}$ are the Fermi velocities of the valley polarized region at valley $\tau$ and moving in the $\alpha$ directions.  Near the zeroth 0-$\pi$ transition point where $m=0$, we find that the CPR from Eq. (\ref{Eq4}) can be approximately written as \cite{NoteX}
\begin{equation}
\label{Eq6}
I_s(\varphi)=I_1\sin(\tilde{\varphi}+\delta)+I_2\sin(2\tilde{\varphi}),
\end{equation}
with $\tilde{\varphi}=\varphi-\Delta_{vp}/E_A$. The coefficients $I_1$, $I_2$ and $\delta$ are
\begin{eqnarray}
 \label{Eq7} I_1 &=& -\frac{16e\cosh\delta_T}{\hbar\beta(1+2\sinh^2\delta_T)}(\frac{2\Delta_{vp}}{E_T}-\frac{\pi}{2}), \\
 \label{Eq8} I_2 &=& \frac{8e\, \text{sech}^2\delta_A}{\hbar\beta(1+2\sinh^2\delta_T)}, \\
 \label{Eq9} \delta &=& -\arctan[\tanh\delta_A\tanh\delta_T/(\frac{2\Delta_{vp}}{E_T}-\frac{\pi}{2})].
\end{eqnarray}
Here, $\delta_A=\pi E_A^{-1}/ \beta$ and $\delta_T=2\pi(\Delta_0^{-1}+E_T^{-1})/\beta$. $E_A=\hbar \delta\bar{v}_{vp}/d$ is the energy scale that reveals the intra-valley inversion breaking, where $\delta\bar{v}_{vp}$ is defined as $\delta\bar{v}_{vp}=1/( v_{vp,++}^{-1}- v_{vp,-+}^{-1})$. For Eq.(\ref{Eq7}), it is clear that $I_1$ changes sign when $\Delta_{vp}$ increases such that $2\Delta_{vp}/E_T>\pi/2$. As a result, the phase of the first harmonic acquires a phase change of $\pi$ which causes the $0$-$\pi$ transition. Interestingly, at higher temperatures, $I_2$ gets suppressed and $I_s$ conforms to a sinusoidal function such that $I_s \approx \sin (\varphi+\phi_0)$. The anomalous phase $\phi_0 =-\Delta_{vp}/E_A$ is the phase shift induced by the valley polarization. More importantly, $\delta$ in Eq.~(\ref{Eq9}), which is the relative phase difference between the first and the second harmonic Josephson currents, induced by the valley polarization and the trigonal warping term, would result in the JDE. As shown in Fig.\ref{fig:fig2}~(d), the analytical results of the Josephson current from Eq.(\ref{Eq6})-(\ref{Eq9}) match the results of the tight binding calculations very well. It is clear from Fig.\ref{fig:fig2}~(d) that there is a large difference between $I_{c+}$ and $I_{c-}$ and thus a large JDE. This is the central result of this work.

Furthermore, the nonreciprocity efficiency $\eta$ as a function of $\Delta_{vp}$, calculated using the 1D lattice model \cite{NoteX} is depicted in Fig.\ref{fig:fig2}~(e). It is interesting to note that $\eta$ is a periodic function of $\Delta_{vp}$ with the same periodicity as the 0-$\pi$ transitions. Near the 0-$\pi$ transitions, $\eta$ is linearly proportional to $\Delta_{vp} - \Delta_{vp}^{m}$, where $\Delta_{vp}^{m} = (m+1/2)\pi E_T/2$. This feature of $\eta$ can be derived from the analytical results of Eq. (\ref{Eq6}).

\begin{figure}
		\centering
		\includegraphics[width=1.0\linewidth]{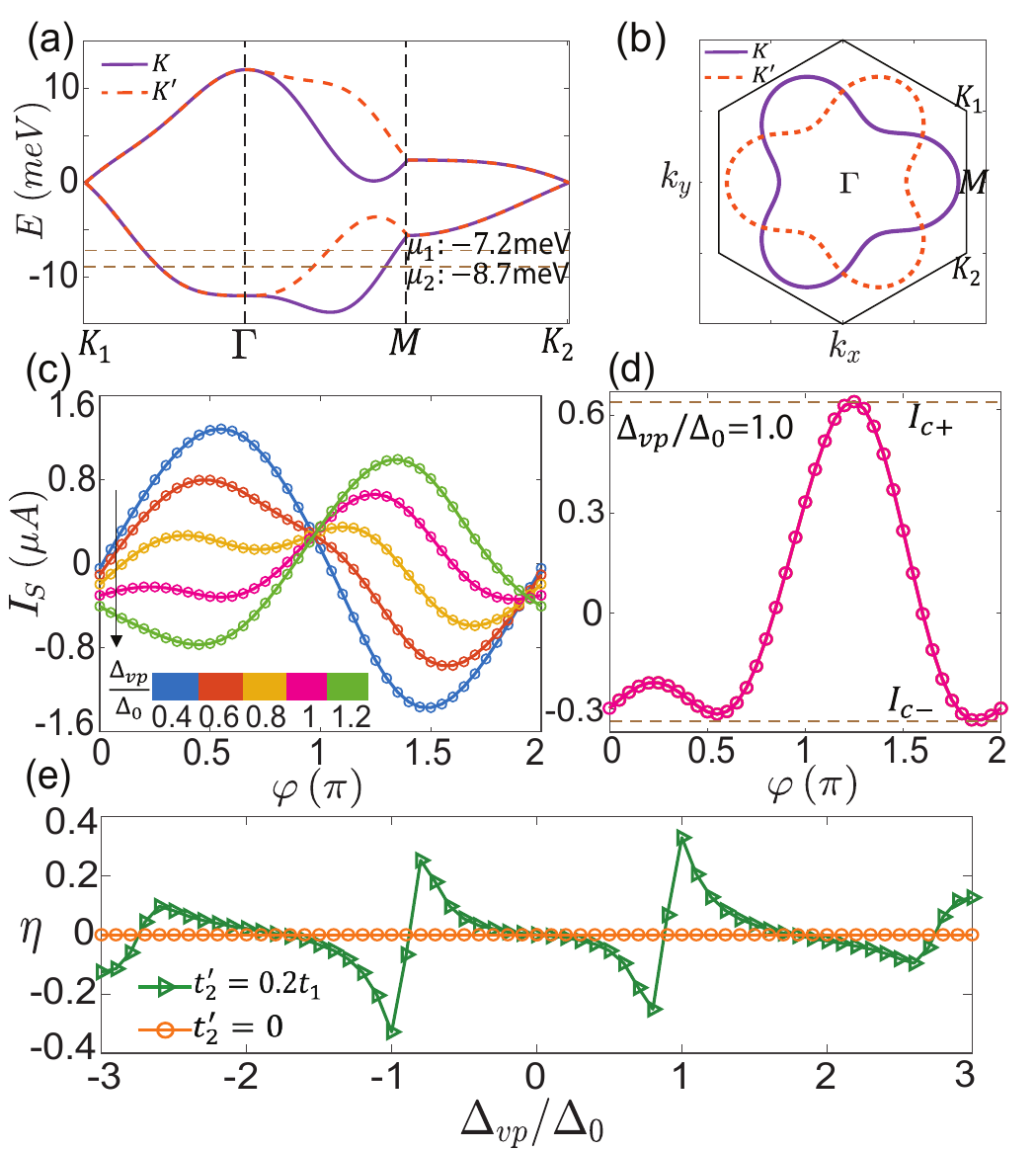}
		\caption{(a) The energy bands of TBG calculated by the tight binding model. (b) The Fermi surfaces of $K$ and $K'$ valleys where the band filling $\nu=-0.5$ [$\mu=-7.2$ meV in (a)]. (c) The Josephson CPR $I_s(\varphi)$ for $\Delta_{vp}/\Delta_0$ going from $0.4$ to $1.2$ and a 0-$\pi$ transition is indicated by the black arrow. (d) The CPR with $\Delta_{vp}/\Delta_0=1$ with $\eta\approx 35\%$. (e) $\eta$ as a function of $\Delta_{vp}/\Delta_0$ with (green triangle) and without (orange circle) warping term $t'_2$. The pairing potential $\Delta_0=1.76k_B T_c\approx 0.4$ meV. The temperature is set to be $k_B T=0.15\Delta_0$. The distance of the weak link is $d=6\sqrt{3}L_M$ and the width of the junction $W_J=25L_M$. The moir\'{e} lattice constant is $L_M \approx 12.8$ nm.}
		\label{fig:fig3}
\end{figure}

\emph{JDE in TBG.}---Experimentally, large JDE was observed in gated-defined JJs with magic-angle TBG \cite{diez2021magnetic} as schematically depicted in Fig.\ref{fig:fig1}~(a). However, the origin of the JDE is not yet known. The JDE was observed only when the weak-link region is gated to half-filling (with two holes per moir\'e unit cell) and it is therefore the property of the weak link. In this section, we extend the 1D model calculations to 2D and show that the valley polarization \cite{po2018origin,lee2019theory,bultinck2020mechanism,liu2021theories} at half-filling combined with trigonally warped Fermi surfaces, naturally give rise to JDE in magic-angle TBG \cite{diez2021magnetic}. To capture the properties of the moir\'e bands in TBG, we use a lattice version of Eq. (\ref{Eq1}) \cite{yuan2018model,koshino2018maximally}. The model can be written as
\begin{equation}
\label{Eq11}
H_0^\tau=\sum_{\braket{ij}}t_1c_{i\tau}^{\dagger}c_{j\tau}+\sum_{\braket{ij}'}(t_2-i\tau t'_2)c^{\dagger}_{i\tau}c_{j\tau}+\text{H.c.}-\sum_{i}\mu_ic^{\dagger}_{i\tau}c_{i\tau}.
\end{equation}
Here $\braket{ij}$ denotes the first nearest hopping terms with amplitude $t_1$, and $\braket{ij}'$ denotes the fifth nearest hopping terms with amplitudes $t_2$ and $t'_2$ respectively. We set $t_2=0.05t_1$, $t'_2=0.2t_1$ in the following calculations. The annihilation operator of an electron with $p_x+i\tau p_y$ orbital on the site $i$ is denoted by  $c_{i\tau}$. For magic-angle TBG, the realistic bandwidth for lowest-energy moir\'{e} bands near charge neutrality is about $20$ meV \cite{pathak2022accurate,carr2019exact}, which corresponds to $t_1=4$ meV. The calculated band structure for the $K$ and $K'$ valleys is shown in Fig.\ref{fig:fig3}~(a). Also, the trigonally warped Fermi surfaces of TBG are shown in Fig.\ref{fig:fig3}~(b). The trigonal warping effect is characterized by $t'_2$ \cite{NoteX}.

In Fig.\ref{fig:fig1}~(a) we show a schematic picture of the gate-defined JJ of TBG with magic-angle TBG. For Gate 1, the filling factor $\nu$ is set to be $\nu=-0.6$, corresponding to the superconducting region [$\mu=-8.7$ meV in Fig.\ref{fig:fig3}~(a)]. For Gate 2, $\nu$ is set to be $\nu=-0.5$, corresponding to the region of valley polarized state [$\mu=-7.2$ meV in Fig.\ref{fig:fig3}~(a)].

By introducing valley polarization $\Delta_{vp}$, we calculate $I_s$ using the lattice Green function approach \cite{NoteX} and the results are shown in Fig.\ref{fig:fig3}~(c). As in the 1D case, there is a 0-$\pi$ transition as $\Delta_{vp}$ increases and the zeroth 0-$\pi$ transition occurs at $\Delta_{vp}/\Delta_0\approx 0.9$. At $\Delta_{vp}/\Delta_0 \approx$ 1, $\eta$ is as large as $35\%$ [Fig.\ref{fig:fig3}~(d)]. Moreover, in Fig.\ref{fig:fig3}~(e) we find that $\eta$ depends sensitively on $\Delta_{vp}$ and has the similar oscillatory behavior as in Fig.\ref{fig:fig2}~(e).  Importantly, we notice that $\eta$ is always zero as the warping term $t'_2$ is turned off [Fig.\ref{fig:fig3}~(e)], which shows that the JDE in TBG further requires the intra-valley inversion symmetry breaking by the trigonal warping effect on top of the time-reversal and inversion symmetry breaking. Interestingly, an additional spin-polarization order parameter can be added to the Hamiltonian and the JDE will only be changed quantitatively as shown in the Supplementary Material \cite{NoteX}.

\emph{Asymmetric Shapiro steps.}---In the sections above, we demonstrate that the unconventional CPR [Eq. (\ref{Eq6})] induced by the valley polarization and the trigonal warping term give rise to JDE. In this section, we propose an alternative method for detecting the unconventional CPR of the SC/VP/SC JJ through the measurement of Shapiro steps \cite{shapiro1963josephson,grimes1968millimeter}. This experiment can be conducted using a resistively shunted Josephson junction (RSJ) model, which is a circuit comprising a JJ in parallel with a resistance R. The current injected into the circuit consists of both the direct current (DC) and the alternating current (AC) components, namely $I(t)=I_{0}+I_{\omega} \cos (\omega t)$, and the DC voltage drop $V_0$ can be measured as shown in Fig.\ref{fig:fig4}~(a). In the RSJ model, the phase dynamics follows \cite{mudi2021model}
\begin{equation}
\label{Eq12}
I_{0}+I_{\omega} \cos (\omega t)=V/R+I_s(\varphi),
\end{equation}
where $V$ is the overall voltage drop on the RSJ, which relates to the phase difference by the second Josephson equation $d \varphi / d t=2 eV/\hbar$. And the DC voltage drop $V_{0}$ on the RSJ is just the time average of $V$, i.e., $V_0=\langle V\rangle_{T}$.

\begin{figure}
		\centering
		\includegraphics[width=1.0\linewidth]{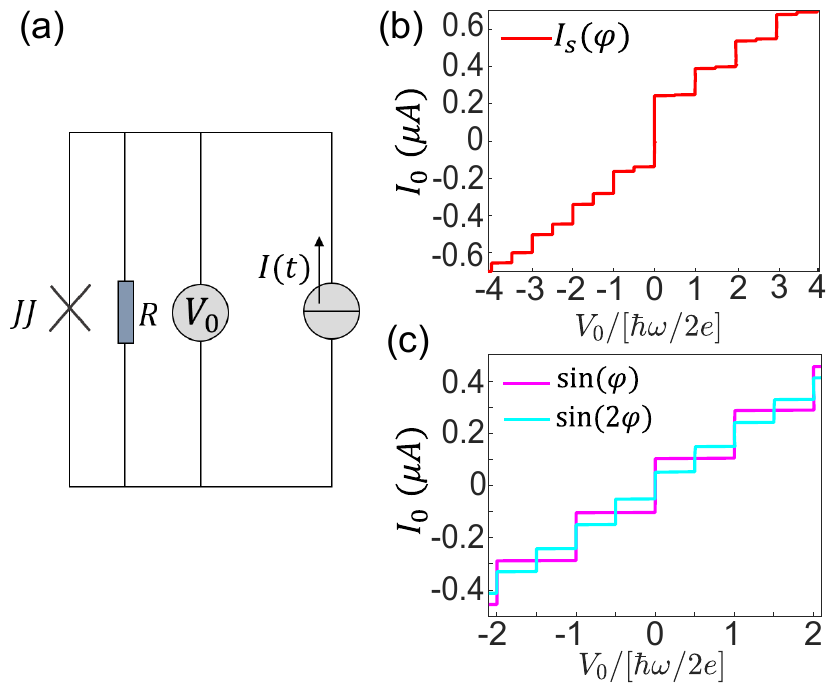}
		\caption{(a) A schematic illustration of the Shapiro steps experiment. The RSJ is driven by the
current $I(t)$, and the DC voltage $V_0$ is measured. (b) and (c) CVC from the model illustrated in (a), with the DC $I_0$ versus the DC voltage drop $V_0$. (b) is the numerical result of CVC, with typical parameters in laboratory $\delta=\pi/3$, $I_{\omega}=0.8 \mu$A, R$=10\Omega$, $\omega=3.14$GHz, $I_1=0.2\mu$A, $I_2=0.8\mu$A. Shapiro steps appear at both integer and half-integer multiples of $\hbar \omega / 2 e$. And an overall asymmetry $I_0(V_0) \ne -I_{0}(-V_0)$ develops as the manifestation of the nonreciprocal nature. (c) Numerical results for $\sin(\varphi)$ and $\sin(2\varphi)$, as comparisons.   }
		\label{fig:fig4}
\end{figure}

We numerically solved the RSJ equation with appropriate parameters for three different kinds of CPRs: $I_s(\varphi)$ as given in Eq. (\ref{Eq6}) with finite $I_1$, $I_2$ and $\delta$, $I_s(\varphi) \propto \sin(\varphi)$, and $I_s(\varphi) \propto \sin(2\varphi)$. The resulting current-voltage characteristics (CVC), in which the DC component $I_0$ as a function of the DC voltage drop $V_0$, are plotted in Fig.\ref{fig:fig4}~(b) and (c). The current jumps of the Shapiro steps occur precisely when the DC voltage matches $V_0=k \hbar \omega / 2 e$, as the integer Shapiro steps, or $V_0= k \hbar \omega / 4 e$, as the half-integer Shapiro steps, where $k=0, \pm 1, \pm 2, \ldots$. We note that near the 0-$\pi$ transition point, the second harmonic component dominates in the CPR, leading to a clear signature of the half-integer Shapiro steps \cite{stoutimore2018second}. Furthermore, compared to $\sin(\varphi)$ and $\sin(2\varphi)$, the CVC of nonreciprocal CPR $I_s(\varphi)$ develops an overall asymmetric character $I_0(V_0)\ne -I_{0}(-V_0)$ at both integer and half-integer Shapiro steps, as the manifestation of the nonreciprocal nature of the junction. A similar asymmetric CVC was also proposed in a SQUID-based circuit very recently \cite{fominov2022asymmetric,souto2022josephson}.

\emph{Discussion.}---Recently, the study of the superconducting diode effect has attracted much attention both experimentally \cite{ando2020observation,lin2022zero} and theoretically \cite{scammell2022theory,daido2022intrinsic,yuan2022supercurrent,he2022phenomenological}. Most of the theories are based on magnetic-field-induced finite-momentum pairings. Some recent theories of JDE also depend on the assumption of finite-momentum pairings in the bulk superconductor induced by magnetic fields. In this work, we show that the JDE can be generated by the weak link of the JJ alone.

We emphasize that the key ingredients for giving rise to the JDE here are the valley polarization and the trigonal warping effect. The pairing symmetries of the superconducting state and the details of the model Hamiltonian are not crucial. For example, it is shown in the Supplementary Material \cite{NoteX} that both the $d$-wave pairing and the $p$-wave pairing support the JDE in TBG. Regarding the model Hamiltonians, a five-band tight binding model of TBG \cite{po2019faithful} is used to calculate the JDE and the results are consistent with the results obtained using Eq. (\ref{Eq11}) \cite{NoteX}.
	
Moreover, although our theory of JDE is based on electron-electron interaction-induced valley polarized states in TBG with trigonal warping terms, our theory can be easily generalized to describe other materials such as rhombohedral trilayer graphene \cite{zhou2021superconductivity,zhou2021half} and Bernal-stacked bilayer graphene \cite{zhou2022isospin,de2022cascade,zhang2022spin} which possess trigonal warping on the Fermi surface. We also expect our theory can apply to spin-polarized systems. In the Supplementary Material \cite{NoteX}, we demonstrate the JDE for a Rashba wire with cubic spin-orbit coupling and an in-plane magnetic field which can be mapped to the valley polarization problem with trigonal warping terms. The model is relevant to recent experiments in which two superconductors are connected by weak links with Rashba spin-orbit coupling and in-plane magnetic fields \cite{pal2022josephson,baumgartner2022supercurrent,jeon2022zero}.

\emph{Acknowledgements.}---We thank Dima Efetov, Jaime Diez, Shuai Chen and Adrian Po for inspiring discussions. K.T.L. acknowledges the support of the Ministry of Science and Technology, China and HKRGC through 2020YFA0309600, RFS2021-6S03, C6025-19G, AoE/P-701/20, 16310520, 16310219, 16307622 and 16309718. Y.M.X. acknowledges the support of HKRGC through PDFS2223-6S01.

\clearpage
		\onecolumngrid
\begin{center}
		\textbf{\large Supplementary Material for\\ ``Valley Polarization Induced Josephson Diode Effect in Twisted Bilayer Graphene''}\\[.2cm]		
      Jin-Xin Hu,$^{1}$  Zi-Ting Sun,$^{1}$  Ying-Ming Xie,$^{1}$  K. T. Law$^{1}$\\[.1cm]
		{\itshape ${}^1$Department of Physics, Hong Kong University of Science and Technology, Clear Water Water Bay,  Hong Kong, China}
\end{center}
	
	\maketitle

\setcounter{equation}{0}
\setcounter{section}{0}
\setcounter{figure}{0}
\setcounter{table}{0}
\setcounter{page}{1}
\renewcommand{\theequation}{S\arabic{equation}}
\renewcommand{\thesection}{ \Roman{section}}

\renewcommand{\thefigure}{S\arabic{figure}}
\renewcommand{\thetable}{\arabic{table}}
\renewcommand{\tablename}{Supplementary Table}

\renewcommand{\bibnumfmt}[1]{[S#1]}
\renewcommand{\citenumfont}[1]{#1}
\makeatletter

\maketitle

\setcounter{equation}{0}
\setcounter{section}{0}
\setcounter{figure}{0}
\setcounter{table}{0}
\setcounter{page}{1}
\renewcommand{\theequation}{S\arabic{equation}}
\renewcommand{\thesection}{ \Roman{section}}

\renewcommand{\thefigure}{S\arabic{figure}}
\renewcommand{\thetable}{\arabic{table}}
\renewcommand{\tablename}{Supplementary Table}

\renewcommand{\bibnumfmt}[1]{[S#1]}
\makeatletter

\maketitle

\section{I. Scattering matrix method for the 1D toy model}
\subsection{A. Model Hamiltonian}
We use a simple continuum model to describe the low-energy physics of a 2D system with valley degrees of freedom. The system possesses trigonally warped Fermi surfaces and interaction-induced valley polarization (VP). The model can be written as
\begin{equation}
H_{\mathrm{eff}}=\lambda_0\left(k_x^2+k_y^2\right)-\mu +\lambda_1 k_x\left(k_x^2-3 k_y^2\right) \tau_z+\Delta_{vp} \tau_z,
\end{equation}
where the $\lambda_0$ term is the kinetic energy, $\mu$ denotes the chemical potential, and $\lambda_1$ is the warping term which is opposite at the two valleys. This Hamiltonian preserves the time-reversal symmetry (TRS) $T=\tau_x K$ and $C_{2y}=\tau_x$ symmetry, but breaks the intra-valley inversion symmetry. The last term is the correlation-induced valley polarization term, which breaks the global TRS.

Now we come to the gate-defined SC/VP/SC Josephson junction, in which the weak-link region is partially valley-polarized. For simplicity, we only consider the 1D special case of the model by fixing $k_y$ as 0, and it is sufficient to capture the essence of the physics. We assume the bandwidth is much larger than the superconductor pairing gap $\Delta_0$ and the valley polarization order parameter $\Delta_{vp}$. Under this condition, we can linearize the dispersion relation in the vicinity of the Fermi energy for a fixed transverse momentum $k_y$ and obtain a low-energy effective Hamiltonian for the Josephson junction
\begin{equation}
H=\frac{1}{2} \sum_{\tau \alpha} \int d x \Psi_{k_y, \tau \alpha}^{\dagger}(x) \hat{H}_{k_y, \tau \alpha}(x) \Psi_{k_y, \tau \alpha}(x).
\end{equation}
Here, $\tau \pm$ labels the $\pm K$ valley, $\alpha=+/-$ labels the right (left) movers near Fermi energy, and $\Psi_{ \tau \alpha}=$ $\left(\psi_{ \tau \alpha}(x), \psi_{-\tau,-\alpha}^{\dagger}(x)\right)^T$ denotes the Nambu basis. The BdG Hamiltonian in terms of this choice of basis reads
\begin{equation}
\hat{H}_{\tau \alpha}(x)=\left(\begin{array}{cc}
-i \hbar v_{f, \tau \alpha}\left( x\right) \partial_x+\tau\Delta_{vp}(x) & \Delta_s(x) \\
\Delta_s(x)^* & -i \hbar v_{f,-\tau-\alpha}\left( x\right) \partial_x+\tau\Delta_{vp}(x)
\end{array}\right),
\end{equation}
with the valley polarization order parameter $\Delta_{vp}(x)=\Delta_{vp}\Theta(x)\Theta(d-x)$ and the pairing potential $\Delta_s(x)=\Delta_0 \left(e^{i \frac{\varphi}{2}} \theta(-x)+e^{-i \frac{\varphi}{2}} \theta(x-d)\right)$, the longitudinal Fermi velocity at a fixed $k_y$ of the superconducting part and junction part is given by $v_{f, \tau \alpha}\left(k_y, x\right)=v_{s, \tau \alpha}\left(k_y\right)[\theta(-x)+\theta(x-d)]+$ $v_{v p, \tau \alpha}\left(k_y\right) \theta(x) \theta(d-x)$ (to make notations more compact, we denote $v_{f, \tau \alpha}\left(k_y, x\right) \equiv v_{f, \tau \alpha}(x)$ below). Here, $\varphi$ is the phase difference, $d$ is the length of the junction, $v_{s, \tau \alpha}, v_{f, \tau \alpha}$ are the longitudinal Fermi momentum along the current direction of the superconducting part and the junction part with valley polarization. One can verify that the whole Hamiltonian $\hat{H}$ (dimension is eight by eight) preserves particle-hole symmetry $P \hat{H} P^{-1}=-\hat{H}$ and breaks TRS: $T \hat{H} T^{-1} \neq \hat{H}$ if $\Delta_{\mathrm{vp}}$ is finite. Here, $\hat{P}=\rho_x \alpha_x \tau_x \hat{K}$, $\hat{T}=\alpha_x \tau_x \hat{K}, \hat{K}$ is complex conjugate, and $\alpha_j$, $\tau_j$, and $\rho_j$ are Pauli matrices defined in $\alpha=+/-$, valley, and particle-hole space, respectively.

\subsection{B. 1-D tight binding model calculations}
In this note, we describe the recursive Green's function method we used to simulate the Josephson junction with details. We simplify the Eq.(S1) to a 1-D model with $k_y=0$, which is also shown in the main text
\begin{equation}
h_\tau=\lambda_0 k_x^2+\tau \lambda_1 k_x^3+\tau \Delta_{vp}-\mu.
\end{equation}
\begin{figure}[h]
		\centering
		\includegraphics[width=0.6\linewidth]{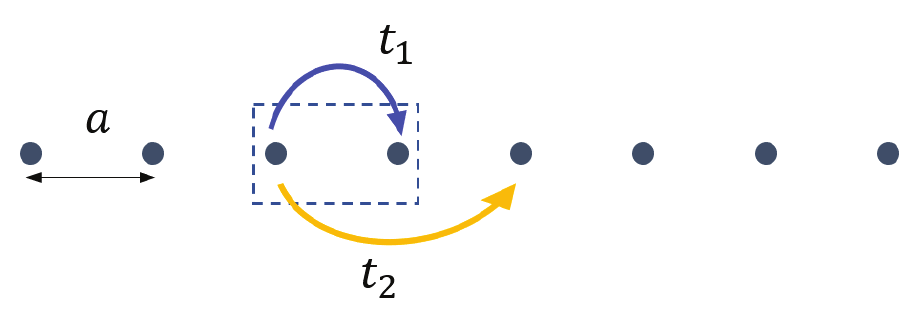}
		\caption{A schematic plot of a 1-D chain, with the first-nearest hopping $t_1$ and the second-nearest hopping $t_2$. We set $a=1$.}
		\label{fig:figS1}
\end{figure}
In Fig.\ref{fig:figS1} we plot the configuration of the 1-D chain for tight binding calculation. The unit cell has two atoms (dashed square) with the first-nearest hopping $t_1$ and the second-nearest hopping $t_2$. In the low energy limit, one can obtain
\begin{equation}
\begin{split}
h_\tau&=\lambda_0 k_x^2+\tau \lambda_1 k_x^3+\tau \Delta_{vp}-\mu \\
&=2\lambda_0(1-\cos k_x)+2\tau\lambda_1 \sin k_x (1-\cos k_x)+\tau \Delta_{vp}-\mu\\
&= 2\lambda_0(1-\cos k_x)+2\tau\lambda_1 \sin k_x -\tau\lambda_1 \sin 2 k_x +\tau \Delta_{vp}-\mu.
\end{split}
\end{equation}
Thus one can obtain $t_1=-\lambda_0-i\tau\lambda_1$ and $t_2=\frac{1}{2}i\tau\lambda_1$.

  In the numerical process, the self-energy $\Sigma_{SC}^{R/L}(i\omega_{n})$
  of the left/right superconducting region is first calculated as
  \begin{equation}
  \Sigma_{SC}^{L/R}(i\omega_{n})=V_{1/N_{x},L/R}^{coup}G_{SC}^{L/R}(i\omega_{n})\left(V_{1/N_{x},L/R}^{coup}\right)^{\dagger}.
  \end{equation}
  Here $G_{SC}^{L/R}(i\omega_{n})$ is the Nambu Green's function of
  the left/right superconducting region with the Matsubara frequency
  $\omega_{n}=(2n+1)\pi k_{B}T$, which can be calculated iteratively
  by assuming the electrodes semi-infinite. $V_{1/N_{x},L/R}^{coup}$
  is the coupling matrix between the left/right end of the central device
  and the corresponding superconducting electrode, the form of which
  can be gotten from $H_{coup}^{R/L}$ in Eq.S5.

  Next we start from the Nambu Green's function of the two ends of the
  central devices
  \begin{align}
  G_{11}^{L}(i\omega_{n}) & =\left[i\omega_{n}-H_{11}^{isol}-\Sigma_{SC}^{L}(i\omega_{n})\right]^{-1},\\
  G_{N_{x}N_{x}}^{R}(i\omega_{n}) & =\left[i\omega_{n}-H_{N_{x}N_{x}}^{isol}-\Sigma_{SC}^{R}(i\omega_{n})\right]^{-1}.
  \end{align}
  Here $H_{xx}^{isol}$ with $x=1,2,...,N_{x}$ represent the BdG Hamiltonian
  of an isolated column (or slice) of the central device.
  Explicitly in our case,
  \begin{equation}
  H_{xx}^{isol}=\left(\begin{array}{cc}
  H_{xx}^{ee}\\
   & H_{xx}^{hh}
  \end{array}\right)
  \end{equation}
  with $H_{xx}^{hh}=-(H_{xx}^{ee})^{*}$. Starting from both ends, the Nambu Green's function of the columns
  (or slices) inside the central device can be calculated recursively
  by projecting Dyson's equation between adjacent columns
  \begin{align}
  \Sigma_{xx}^{L}(i\omega_{n}) & =V_{x+1,x}G_{xx}^{L}(i\omega_{n})V_{x+1,x}^{\dagger}\\
  \Sigma_{xx}^{R}(i\omega_{n}) & =V_{x-1,x}G_{xx}^{R}(i\omega_{n})V_{x-1,x}^{\dagger}\\
  G_{x+1,x+1}^{L}(i\omega_{n}) & =\left[i\omega_{n}-H_{x+1,x+1}^{isol}-\Sigma_{xx}^{L}(i\omega_{n})\right]^{-1}\\
  G_{x-1,x-1}^{R}(i\omega_{n}) & =\left[i\omega_{n}-H_{x-1,x-1}^{isol}-\Sigma_{xx}^{R}(i\omega_{n})\right]^{-1}.
  \end{align}
  Explicitly in our case
  \[
  V_{x+1,x}^{\dagger}=V_{x-1,x}=\left(\begin{array}{cc}
  V_{x}^{ee}\\
   & V_{x}^{hh}
  \end{array}\right)
  \]
  where $V_{x}^{hh}=-\left(V_{x}^{ee}\right)^{*}$. Note that the superscript
  $R/L$ means that the Nambu Green's function $G^{R/L}$ here only
  represents the right/left part of the device. We need to glue them
  together to get the Nambu Green's function of the whole device
  \begin{equation}
  G_{xx}(i\omega_{n})=\left[i\omega_{n}-H_{x,x}^{isol}-\Sigma_{x-1,x-1}^{L}(i\omega_{n})-\Sigma_{x+1,x+1}^{R}(i\omega_{n})\right]^{-1}.
  \end{equation}
  Furthermore, we can also get
  \begin{align}
  G_{x+1,x}(i\omega_{n}) & =G_{x+1,x+1}^{R}(i\omega_{n})V_{x+1,x}G_{x,x}(i\omega_{n}),\\
  G_{x,x+1}(i\omega_{n}) & =G_{xx}(i\omega_{n})V_{x,x+1}G_{x+1,x+1}^{R}(i\omega_{n}).
  \end{align}
  And the Josephson current can be calculated as
  \begin{equation}
  I_{S}(\varphi)=\frac{2ek_{B}T}{\hbar}\mathrm{Im}\sum_{\omega_{n}}\mathrm{Tr}\left[\tilde{V}_{x,x+1}G_{x+1,x}(i\omega_{n})-\tilde{V}_{x,x+1}^{\dagger}G_{x,x+1}(i\omega_{n})\right]
  \end{equation}
  with $\tilde{V}_{x,x+1}=\left(\begin{array}{cc}
  V_{x}^{ee}\\
   & -V_{x}^{hh}
  \end{array}\right)=\left(\begin{array}{cc}
  V_{x}^{ee}\\
   & (V_{x}^{ee})^{*}
  \end{array}\right)$.
\subsection{C. Scattering matrix method}
The scattering eigenstates in the SC region of the left $(\mathrm{L})$ and right $(\mathrm{R})$ sides can be written as:
\begin{equation}
\begin{aligned}
&\psi_{s, \tau \alpha}^L=\left(\begin{array}{c}
e^{-i \alpha \beta} \\
e^{-i \frac{\varphi}{2}}
\end{array}\right) e^{i k_{s, \tau \alpha}^0 x+\kappa_{\tau \alpha} x}, x \leq 0, \\
&\psi_{s, \tau \alpha}^R=\left(\begin{array}{c}
e^{i \alpha \beta} \\
e^{i \frac{\varphi}{2}}
\end{array}\right) e^{i k_{s, \tau \alpha}^0(x-d)-\kappa_{\tau \alpha}(x-d)}, x \geq d,
\end{aligned}
\end{equation}
with $k_{s, \tau \alpha}^0$ being the Fermi momentum and $\kappa_{\tau \alpha}$, $\beta$ are given by:
\begin{equation}
\begin{aligned}
\kappa_{\tau \alpha} &=\frac{\sqrt{\Delta_{\mathrm{s}}^2-\epsilon^2}}{\alpha\hbar v_{s, \tau \alpha}}, \\
\beta &=\left\{\begin{array}{l}
\operatorname{acos} \frac{\epsilon}{\Delta_{\mathrm{s}}}, \text { if } \epsilon<\Delta_{\mathrm{s}} \\
-i \operatorname{acosh} \frac{\epsilon}{\Delta_{\mathrm{s}}}, \text { if } \epsilon>\Delta_{\mathrm{s}}.
\end{array}\right.
\end{aligned}
\end{equation}

The scattering states in the VP region $(0 \leq x \leq d)$ can be written as:
\begin{equation}
\begin{aligned}
&\psi_{v p, e, \tau \alpha}=\frac{1}{\sqrt{N_{e, \tau \alpha}}}\left(\begin{array}{l}
1 \\
0
\end{array}\right) e^{i k_{e, \tau \alpha} x} \\
&\psi_{v p, h, \tau \alpha}=\frac{1}{\sqrt{N_{h, \tau \alpha}}}\left(\begin{array}{l}
0 \\
1
\end{array}\right) e^{i k_{h, \tau \alpha} x} .
\end{aligned}
\end{equation}
Here, $k_{e, \tau \alpha}$ and $k_{h, \tau \alpha}$ are the wave vectors for electron and hole states, respectively, and $N_{e(h), \tau \alpha}$ are normalization factors to ensure that the scattering matrices are unitary. Here, $\psi_{v p, e, \tau+}, \psi_{v p, h, \tau-}$ are right movers, while $\psi_{v p, e, \tau-}, \psi_{v p, h, \tau+}$ are left movers. If we only consider the linear order correction of Fermi momenta due to the VP, the correction can be written as $k_{e, \tau \alpha} \approx k_{v p, \tau \alpha}^0+\delta k_{e, \tau \alpha}, k_{h, \tau \alpha} \approx  k_{v p, \tau \alpha}^0+\delta k_{h, \tau \alpha}$ in which
\begin{equation}
\begin{array}{r}
\delta k_{e, \tau \alpha}=\frac{\epsilon-\tau \Delta_{\mathrm{vp}}}{ \hbar v_{v p, \tau \alpha}}, \\
\delta k_{h, \tau \alpha}=\frac{\epsilon-\tau \Delta_{\mathrm{vp}}}{\hbar v_{v p,-\tau-\alpha}} .
\end{array}
\end{equation}

This approximation will be used in the calculation of current-phase relation (CPR) later.

As the BdG Hamiltonian is block-diagonalized, we can solve the scattering matrix for $\tau=\pm $ separately, then add their contributions to the Josephson current together. We take the wave function of the whole junction as
\begin{equation}
\psi_{\tau}(x)=\left\{
    \begin{aligned}
    &a \psi_{s, \tau+}^L(x)+b \psi_{s,\tau-}^L(x)     & \text { if } x \leq  0   \\
    &c_e^{+} \psi_{v p, e, \tau+}(x)+c_e^{-} \psi_{v p, e, \tau-}(x)+c_h^{+} \psi_{v p, h, \tau+}(x)+c_h^{-} \psi_{v p, h, \tau-}(x)     &\text { if }0 \leq  x \leq d   \\
    &a^{\prime} \psi_{s, \tau+}^R(x)+b^{\prime} \psi_{s, \tau-}^R(x)     &\text { if } x \geq d
    \end{aligned}
\right.
\end{equation}

In the case of perfectly transparent contacts, the boundary conditions at $x=0$ and $x=d$ read
\begin{equation}
\begin{aligned}
&a \psi_{s, \tau+}^L(x=0)+b \psi_{s, \tau-}^L(x=0)=c_e^{+} \psi_{v p, e, \tau+}(x=0)+c_e^{-} \psi_{v p, e, \tau-}(x=0) \\
&+c_h^{+} \psi_{v p, h, \tau+}(x=0)+c_h^{-} \psi_{v p, h, \tau-}(x=0) \\
&a^{\prime} \psi_{s, \tau+}^R(x=d)+b^{\prime} \psi_{s, \tau-}^R(x=d)=c_e^{+} \psi_{v p, e, \tau+}(x=d)+c_e^{-} \psi_{v p, e, \tau-}(x=d) \\
&+c_h^{+} \psi_{v p, h, \tau+}(x=d)+c_h^{-} \psi_{v p, h, \tau-}(x=d) \\
&a v_{s, \tau+} \psi_{s, \tau+}^L(x=0)+b v_{s, \tau-} \psi_{s, \tau-}^L(x=0)=v_{v p, \tau+} c_e^{+} \psi_{v p, e, \tau+}(x=0)+v_{v p, \tau-} c_e^{-} \psi_{v p, e, \tau-}(x=0) \\
&-v_{v p,-\tau-} c_h^{+} \psi_{v p, h, \tau+}(x=0)-v_{v p,-\tau+} c_h^{-} \psi_{v p, h, \tau-}(x=0) \\
&a^{\prime} v_{s, \tau+} k_{0, \tau+} \psi_{s, \tau+}^R(x=d)+b^{\prime} v_{s, \tau-} \psi_{s, \tau-}^R(x=d)=v_{v p, \tau+} c_e^{+} \psi_{v p, e, \tau+}(x=d)+v_{v p, \tau-} c_e^{-} \psi_{v p, e, \tau-}(x=d) \\
&-v_{v p,-\tau-} c_h^{+} \psi_{v p, h, \tau+}(x=d)-v_{v p,-\tau+} c_h^{-} \psi_{v p, h, \tau-}(x=d).
\end{aligned}
\end{equation}
Here, the first and second equalities are obtained from the continuity of the whole wave function, while the third and fourth equalities are obtained from the conservation of particle current.

 For the convenience of the later discussions, we define
\begin{equation}
\begin{array}{r}
a(L)=a, b(L)=b, a(R)=a^{\prime}, b(R)=b^{\prime} \\
c_e^{\dagger}(L)=c_e^{+}, c_e^{-}(L)=c_e^{-}, c_h^{+}(L)=c_h^{+}, c_h^{-}(L)=c_h^{-} \\
c_e^{\dagger}(R)=c_e^{+} e^{ik_{e, \tau+} d}, c_e^{-}(R)=c_e^{-} e^{ik_{e, \tau-} d} \\
c_h^{+}(R)=c_h^{+} e^{ik_{h, \tau+} d}, c_h^{-}(R)=c_h^{-} e^{ik_{h, \tau-} d}.
\end{array}
\end{equation}

Now, we can write down both the scattering matrices describing the Andreev scatterings at each interface and the ones in the normal region. For simplicity, we make the Andreev approximation that there is no chemical potential difference at the SC-VP interface. In this case, $v_{v p, \tau \pm}=v_{s, \tau \pm}$, and the normalization factors in the scattering states of VP region can be simply taken as $N_{e, \tau \alpha}=N_{h, \tau \alpha}=1$. And in the normal region, we only consider one channel and the clean limit. By solving the linear equations, we obtain the S-matrices which can be written as:

\begin{equation}
\psi_{\text {out }}=\left(\begin{array}{c}
c_e^{+}(L) \\
c_h^{-}(L) \\
c_e^{-}(R) \\
c_h^{+}(R)
\end{array}\right)=\left(\begin{array}{cccc}
0 & e^{ i \frac{\phi}{2}-i \beta} & 0 & 0 \\
e^{- i \frac{\phi}{2}-i \beta} & 0 & 0 & 0 \\
0 & 0 & 0 & e^{- i \frac{\phi}{2}-i \beta} \\
0 & 0 & e^{ i \frac{\phi}{2}-i \beta} & 0
\end{array}\right)\left(\begin{array}{c}
c_e^{-}(L) \\
c_h^{+}(L) \\
c_e^{+}(R) \\
c_h^{-}(R)
\end{array}\right) \equiv \mathcal{S}_{\mathrm{A}} \psi_{in},
\end{equation}
\begin{equation}
\psi_{\text {in }}=\left(\begin{array}{l}
c_e^{-}(L) \\
c_h^{+}(L) \\
c_e^{+}(R) \\
c_h^{-}(R)
\end{array}\right)=\left(\begin{array}{cccc}
0 & 0 & e^{-ik_{e, \tau-} d} & 0 \\
0 & 0 & 0 & e^{-ik_{h, \tau+} d} \\
e^{ik_{e, \tau+} d} & 0 & 0 & 0 \\
0 & e^{ik_{h, \tau-} d} & 0 & 0
\end{array}\right)\left(\begin{array}{l}
c_e^{+}(L) \\
c_h^{-}(L) \\
c_e^{-}(R) \\
c_h^{+}(R)
\end{array}\right) \equiv \mathcal{S}_{\mathrm{N}} \psi_{out}.
\end{equation}
Here, we note that only the amplitudes of Andreev reflections in $\mathcal{S}_{\mathrm{A}}$ are finite due to the absence of momentum mismatches. We also define the energy scales with $E_T=\hbar \bar{v}_{vp}/d$, and $\bar{v}_{vp}=2/( v_{vp,++}^{-1}+ v_{vp,-+}^{-1})$ which is the Thouless energy and $E_A=\hbar \delta\bar{v}_{vp}/d$ is the energy scale that reveals the intra-valley inversion breaking, where $\delta\bar{v}_{vp}$ is defined as $\delta\bar{v}_{vp}=1/( v_{vp,++}^{-1}- v_{vp,-+}^{-1})$.

\subsection{D. Calculations of the Josephson current}

We calculate the Josephson current for the 1D model using the scattering matrix method\cite{brouwer1997anomalous}

\begin{equation}
\begin{split}
I_s(\varphi)&=-\frac{4e}{\hbar\beta}\frac{d}{d\varphi}\sum_{n=0}^{\infty}\ln \text{det}[1-S_A(i\omega_n,\varphi)S_N(i\omega_n,\varphi)]\\
&=-\frac{4e}{\hbar\beta}\frac{d}{d\varphi}\sum_{n=0}^{\infty}\sum_{\tau=\pm}\ln [\cos(2\beta-\frac{2(i\omega_n-\tau\Delta_{vp})}{E_T})-\cos(\varphi+\frac{i\omega_n-\tau\Delta_{vp}}{\tau E_A})]\\
&=-\frac{4e}{\hbar\beta}\sum_{n=0}^{\infty}\sum_{\tau=\pm}\frac{\sin(\varphi+\frac{i\omega_n-\tau\Delta_{vp}}{\tau E_A})}{\cos(2\beta-\frac{2(i\omega_n-\tau\Delta_{vp})}{E_T})-\cos(\varphi+\frac{i\omega_n-\tau\Delta_{vp}}{\tau E_A})}\\
&=-\frac{4e}{\hbar\beta}\sum_{n=0}^{\infty}\sum_{\tau=\pm}\frac{\sin(\tilde{\varphi}+\frac{i\tau\omega_n}{E_A})}{\cos(2\beta-\frac{2(i\omega_n-\tau\Delta_{vp})}{E_T})-\cos(\tilde{\varphi}+\frac{i\tau\omega_n}{E_A})}.
\end{split}
\end{equation}
Here, $\tilde{\varphi}=\varphi-\Delta_{vp}/E_A$. We make some assumptions to evaluate the summation. First, the Matsubara frequency $\omega=(2n+1)\pi/\beta$ decreases quickly with $n$, so we can only maintain the $n=0$ term. Second, since the two valleys are related by time-reversal symmetry, one can check that the summations for the two valleys are complex conjugate with each other. Thus the Josephson current can be evaluated as

\begin{equation}
\begin{split}
I_s(\varphi)&=-\frac{8e}{\hbar\beta}\text{Re}[\frac{\sin(\tilde{\varphi}+\frac{i\pi T}{E_A})}{\cos(2\beta-\frac{2(i\pi T-\Delta_{vp})}{E_T})-\cos(\tilde{\varphi}+\frac{i\pi T}{E_A})}]\\
&= -\frac{8e}{\hbar\beta}\text{Re}[\frac{\sin\tilde{\varphi}+i\tanh(\delta_A)\cos\tilde{\varphi}}{(-\cos \frac{2\Delta_{vp}}{E_T}\cosh\delta_T-\cos\tilde{\varphi})+i(\sin\tilde{\varphi}\tanh\delta_A-\sin \frac{2\Delta_{vp}}{E_T}\sinh\delta_T)}]\\
&=-\frac{8e}{\hbar\beta}\frac{-\frac{1}{2}\text{sech}^2\delta_A\sin 2\tilde{\varphi}-\cos \frac{2\Delta_{vp}}{E_T}\cosh\delta_T\sin\tilde{\varphi}-2\sinh\delta_T\tanh\delta_A\cos\tilde{\varphi}}{(\cos \frac{2\Delta_{vp}}{E_T}\cosh\delta_T+\cos\tilde{\varphi})^2+(\sin\tilde{\varphi}\tanh\delta_A-\sin \frac{2\Delta_{vp}}{E_T}\sinh\delta_T)^2}.
\end{split}
\end{equation}
It is clear that 0-$\pi$ transitions occur at $2\Delta_{vp}/E_T=(n+\frac{1}{2})\pi$. Expand $J_s(\varphi)$ near $2\Delta_{vp}/E_T=\pi/2$, one can obtain

\begin{equation}
I_s(\varphi)=\frac{8e}{\hbar\beta(1+2\sinh^2\delta_T)}[\text{sech}^2\delta_A\sin 2\tilde{\varphi}-2(\frac{2\Delta_{vp}}{E_T}-\frac{\pi}{2})\cosh\delta_T\sin(\tilde{\varphi}+\delta)],
\end{equation}
with
\begin{equation}
\delta = -\arctan[\tanh\delta_A\tanh\delta_T/(\frac{2\Delta_{vp}}{E_T}-\frac{\pi}{2})].
\end{equation}
In the calculations we use some approximations: First, we treat $\delta_A$ is a small comparable to $\delta_T$. Second, we use $\sin(x)/(\cos^2 x+a)\approx 2\sin(x)/(1+2a)$.
\subsection{E. Effect of spin polarization}
Now we add the spin degree of freedom by considering  interaction induced spin polarization(SP). We rewrite the BDG Hamiltonian in the basis
$\Psi_{\tau s \alpha}=$ $\left(\psi_{ \tau s \alpha}(x), \psi_{-\tau,-s,-\alpha}^{\dagger}(x)\right)^T$ denotes the Nambu basis. The BdG Hamiltonian in terms of this choice of basis reads
\begin{equation}
\hat{H}_{\tau s \alpha}(x)=\left(\begin{array}{cc}
-i \hbar v_{f, \tau s \alpha}\left( x\right) \partial_x+\tau\Delta_{vp}(x)+s\Delta_{sp}(x)  & \Delta_{\mathrm{s}}(x) \\
\Delta_{\mathrm{s}}^*(x) & -i \hbar v_{f,-\tau -s -\alpha}\left( x\right) \partial_x+\tau\Delta_{vp}(x)+s\Delta_{sp}(x)
\end{array}\right).
\end{equation}
Follow the same procedures of Eq.(S15),Eq.(S16), one can obtain
\begin{equation}
J_s(\varphi)=\frac{8e}{\hbar\beta(1+2\sinh^2\delta_T)}\sum_{s=\pm}[sech^2\delta_A\sin 2(\tilde{\varphi}-s\frac{\Delta_{sp}}{E_A})-2(\frac{2\Delta_{vp}+2s\Delta_{sp}}{E_T}-\frac{\pi}{2})\cosh\delta_T\sin(\tilde{\varphi}-s\frac{\Delta_{sp}}{E_A}+\delta_s)]
\end{equation}
with
\begin{equation}
\delta_s = -\arctan[\tanh\delta_A\tanh\delta_T/(\frac{2\Delta_{vp}+2s\Delta_{sp}}{E_T}-\frac{\pi}{2})].
\end{equation}
For each mode of $s=\pm$, the $0-\pi$ transition has a shift by $\Delta_{sp}$. We will study the effect of spin polarization in the future.

\section{II. Josephson diode effect in TBG with unconventional pairings}
\begin{figure}[h]
	\centering
	\includegraphics[width=1
	\linewidth]{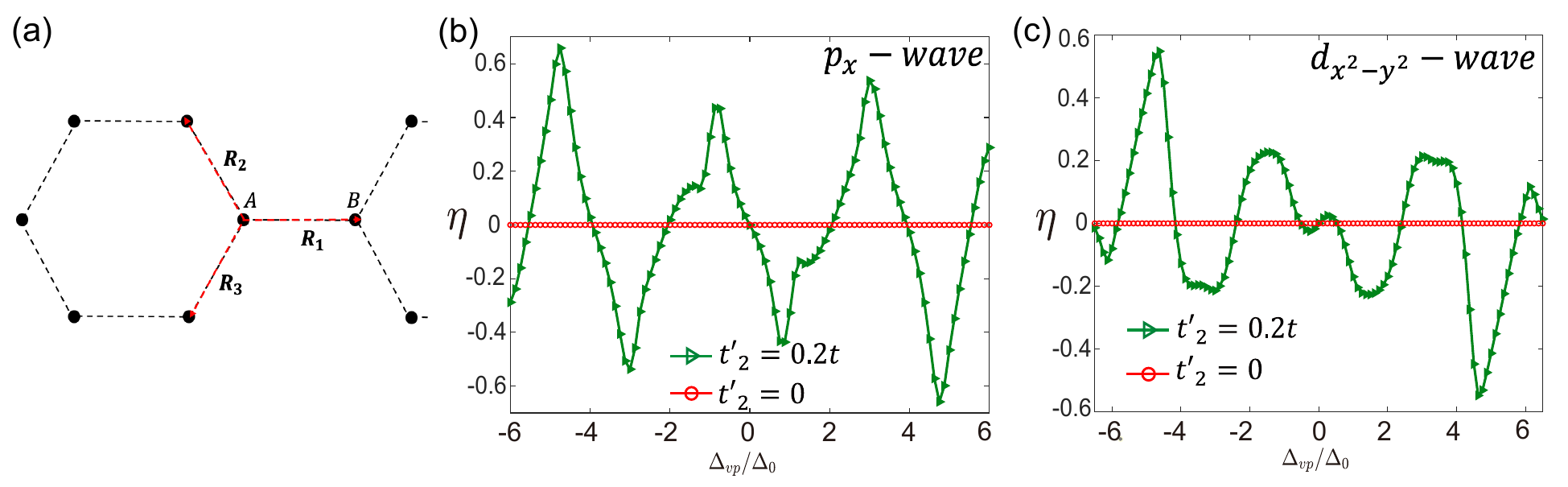}
	\caption{(a) \addYM{A schematic diagram of the honeycomb lattice}. The two sublattices are denoted by the A and B \addYM{black dots}. The nearest-neighbor vectors $\bm{R_{1,2,3}}$ are shown as the red arrows. (b) and (c) \addYM{the diode efficiency} $\eta$ as a function of $\Delta_{vp}/\Delta_0$ with (green triangle) and without (red circle) warping term $t'_2$ for $p_x$\addYM{-wave} and $d_{x^2-y^2}$\addYM{-wave} pairing respectively. The pairing potential $\Delta_0=0.4$ meV. The temperature is set to be $k_B T=0.15\Delta_0$.}
	\label{fig:figS2}
\end{figure}

In this section, motivated by the recent experiment exhibiting the possibilities of unconventional pairing in TBG \cite{oh2021evidence}, we study the Josephson diode effect in TBG with unconventional pairing. In their experiment, the low-energy region of the V-shaped gap supports an anisotropic pairing mechanism with nodes in the superconducting gap function. The STS spectra resembles the quasiparticle DOS of a nodal superconductor, as for higher-angular-momentum pairing (such as $p$- or $d$-wave pairing) with an anisotropic gap function, which was theoretically studied before \cite{wu2018theory,wu2019identification}. To our knowledge, the existing experimental data is not sufficient to distinguish the spin-triplet $p$-wave (odd orbital parity) and spin-singlet $d$-wave (even orbital parity). For the $p$-wave pairing, the pairing potential are $\Delta^p_{x}$ and $\Delta^p_{y}$, which are two degenerate functions in $E_{1}$ representation of $D_6$ point group. For the $d$-wave pairing, the pairing potential are $\Delta^d_{x^2-y^2}$ and $\Delta^d_{xy}$, which are two degenerate functions in $E_{2}$ representation of $D_6$ point group. In a honeycomb lattice ([Fig.\ref{fig:figS2}(a)]), they can be written as \cite{brydon2019loop}
\begin{equation}
	\begin{split}
		& \Delta^p_{x}(\bm{k}) =i \Delta_0 \{[\cos(k_x a)-\cos(\frac{1}{2}k_x a)\cos(\frac{\sqrt{3}}{2}k_y a)]s_y+[\sin(k_x a)+\sin(\frac{1}{2}k_x a)\cos(\frac{\sqrt{3}}{2}k_y a)]s_x\},\\
		& \Delta^p_{y}(\bm{k}) = i \sqrt{3}\Delta_0\{\sin(\frac{1}{2}k_x a)\sin(\frac{\sqrt{3}}{2}k_y a)s_y+\cos(\frac{1}{2}k_x a)\sin(\frac{\sqrt{3}}{2}k_y a)s_x\}
	\end{split}
\end{equation}
and
\begin{equation}
	\begin{split}
		& \Delta^d_{x^2-y^2}(\bm{k}) = \Delta_0 \{[\cos(k_x a)-\cos(\frac{1}{2}k_x a)\cos(\frac{\sqrt{3}}{2}k_y a)]s_x-[\sin(k_x a)+\sin(\frac{1}{2}k_x a)\cos(\frac{\sqrt{3}}{2}k_y a)]s_y\},\\
		& \Delta^d_{xy}(\bm{k}) = \sqrt{3}\Delta_0\{ -\sin(\frac{1}{2}k_x a)\sin(\frac{\sqrt{3}}{2}k_y a)s_x+\cos(\frac{1}{2}k_x a)\sin(\frac{\sqrt{3}}{2}k_y a)s_y \}.
	\end{split}
\end{equation}
Here, $\Delta_0$ is the pairing potential, $a$ is the nearest bond distance that is $1/\sqrt{3}$ times the moir\'{e} lattice constant $L_M$, $s_x$ and $s_y$ are Pauli matrices defined in the AB sublattice space.
For simplicity, we consider the case of $p_x$ and $d_{x^2-y^2}$ pairing in TBG. More explicitly, the pairing function $\Delta^p_{x}(\bm{k})$ and $\Delta^d_{x^2-y^2}(\bm{k})$ can be written on the lattice bond [see Fig.\ref{fig:figS2}(a)], which reads
\begin{equation}
	\Delta^p_x(\bm{k})=\Delta_0 \sum_{j=1}^3 \cos(\phi_j)
	\left(\begin{array}{cc}
		0 & e^{i\bm{k}\cdot \bm{R}_j}\\
		-e^{-i\bm{k}\cdot \bm{R}_j} & 0
	\end{array}\right),
	\Delta^d_{x^2-y^2}(\bm{k})=\Delta_0 \sum_{j=1}^3 \cos(\phi_j)
	\left(\begin{array}{cc}
		0 & e^{i\bm{k}\cdot \bm{R}_j}\\
		e^{-i\bm{k}\cdot \bm{R}_j} & 0
	\end{array}\right).
\end{equation}
Here, the phase for the bond $j$ is given by $\phi_j=2(j-1)\pi/3$. For example, the Bogoliubov–de Gennes (BdG) Hamiltonian of TBG with $d_{x^2-y^2}$ pairing is given by
\begin{equation}
	H_{BdG}(\bm{k})=\left(\begin{array}{cc}
		H_0^{\tau=+}(\bm{k}) & \Delta^d_{x^2-y^2}(\bm{k}) \\
		\Delta_{x^2-y^2}^{d \dagger}(\bm{k}) & \addYM{-H_0^{\tau=-*} (-\bm{k})}
	\end{array}\right),
\end{equation}

where the noraml $H_0^{\tau}(\bm{k})$ is given by the tight binding model Eq.~(10) in the main text.

  In the tight binding model, the $p$ or $d$ wave pairing is added in the nearest hopping between A and B site, while the conventional s wave pairing is added in the intra-site. In Fig.\ref{fig:figS2}(b) and (c) we plot the calculated diode efficiency $\eta$ as a function of $\Delta_{vp}$ for $p_x$\addYM{-wave} and $d_{x^2-y^2}$\addYM{-wave} pairing respectively. Importantly, $\eta$ is always zero \addYM{when} the warping term $t'_2$ is turned off. Thus \addYM{the JDE identified by us  can also be applied to unconventional  pairing functions}. Interestingly, $\eta$ can reach up to $60\%$ with $p$\addYM{-} and $d$\addYM{-}wave pairing, which is \addYM{even} larger than that with s\addYM{-}wave pairing. This is because apart from the second harmonic term, higher order scattering processes may occur in the case of anisotropic pairing. \addYM{However}, due to the anisotropy of the pairing, \addYM{we expect} the details of the diode effect would also depend on the orientation of the pairings.

\section{III. Tight binding models for twisted bilayer graphene}
\begin{figure}[h]
	\centering
	\includegraphics[width=1
	\linewidth]{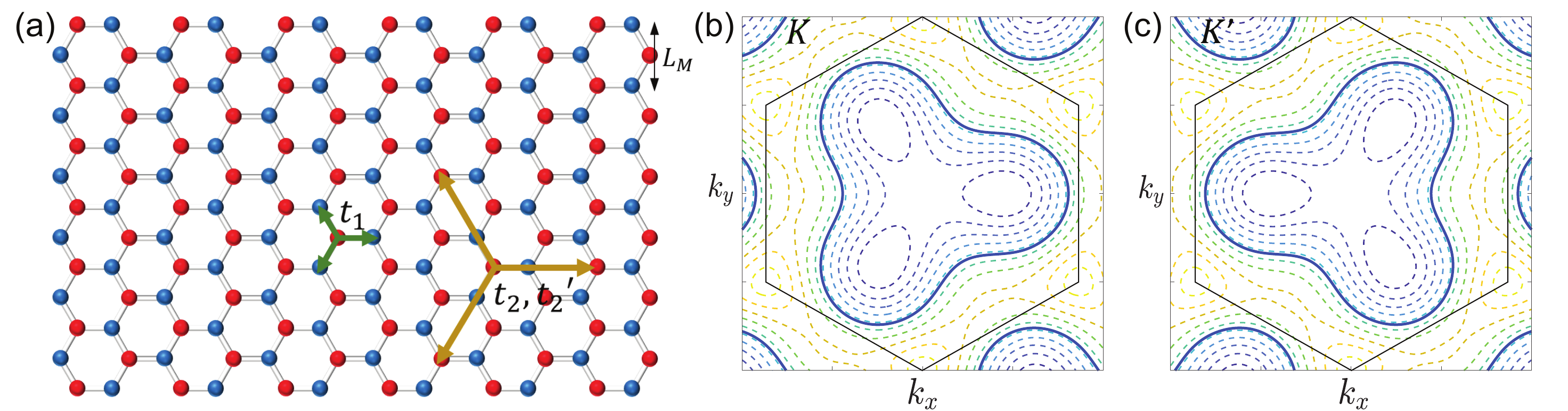}
	\caption{(a) A schematic plot of a honeycomb lattice. The first-nearest and fifth-nearest hopping are labelled by $t_1,t_2,t_2'$. (b) ((c))The Fermi surface of TBG for the $K$ ($K'$) valley. The Fermi circles for filling $\nu=-0.5$ are labelled by thick solid lines.}
	\label{fig:figS3}
\end{figure}

\subsection{A. Minimal two-band model}
Here, the tight binding model for twisted bilayer graphene used for the calculations of the main text is described in more details. With the basis $(c_{A\tau}(\bm{k}),c_{B\tau}(\bm{k}))^T$, we write the tight binding model as
\begin{equation}
	H^\tau(\bm{k})=\left(
	\begin{matrix}{}
		&2t_2 h_2+2\tau t_2' h_2'  & t_1 h_1   \\
		&t_1 h_1^*  &  2t_2 h_2+2\tau t_2' h_2'
	\end{matrix}\right),
\end{equation}
with
\begin{equation}
	h_1=e^{ik_x/\sqrt{3}}+e^{i(-k_x/2+\sqrt{3}k_y/2)/\sqrt{3}}+e^{i(-k_x/2-\sqrt{3}k_y/2)/\sqrt{3}}
\end{equation}
\begin{equation}
	h_2=\cos(\sqrt{3}k_x)+\cos(\sqrt{3}(-k_x/2-\sqrt{3}k_y/2))+\cos(\sqrt{3}(-k_x/2+\sqrt{3}k_y/2))
\end{equation}
\begin{equation}
	h_2'=\sin(\sqrt{3}k_x)+\sin(\sqrt{3}(-k_x/2-\sqrt{3}k_y/2))+\sin(\sqrt{3}(-k_x/2+\sqrt{3}k_y/2)).
\end{equation}
Here $k_x,k_y$ are in the units of $L_M^{-1}$. In Fig.\ref{fig:figS3}~(a) we plot the hopping term $t_1,t_2,t_2'$. By expanding $H^\tau(\bm{k})$ at $\Gamma$ point, one can obtain the energy dispersion of the hole band:
\begin{equation}
	E^\tau(\bm{k})=(-3+\frac{1}{4}\bm{k}^2)t_1+(6-\frac{9}{2}\bm{k}^2)t_2-\frac{3\sqrt{3}}{4}\tau t_2'(k_x^3-3k_x k_y^2).
\end{equation}
Comparing with Eq.(S1), one can obtain $\lambda_0=\frac{1}{4}t_1-\frac{9}{2}t_2$, $\lambda_1=-\frac{3\sqrt{3}}{4}t_2'$. In Fig.\ref{fig:figS3}~(b),(c) the trigonally warped Fermi surfaces are shown. It is clear that the trigonal warping effect is determined by $t'_2$.

\subsection{B. Faithful five-band model}

\begin{figure}[h]
	\centering
	\includegraphics[width=0.8\linewidth]{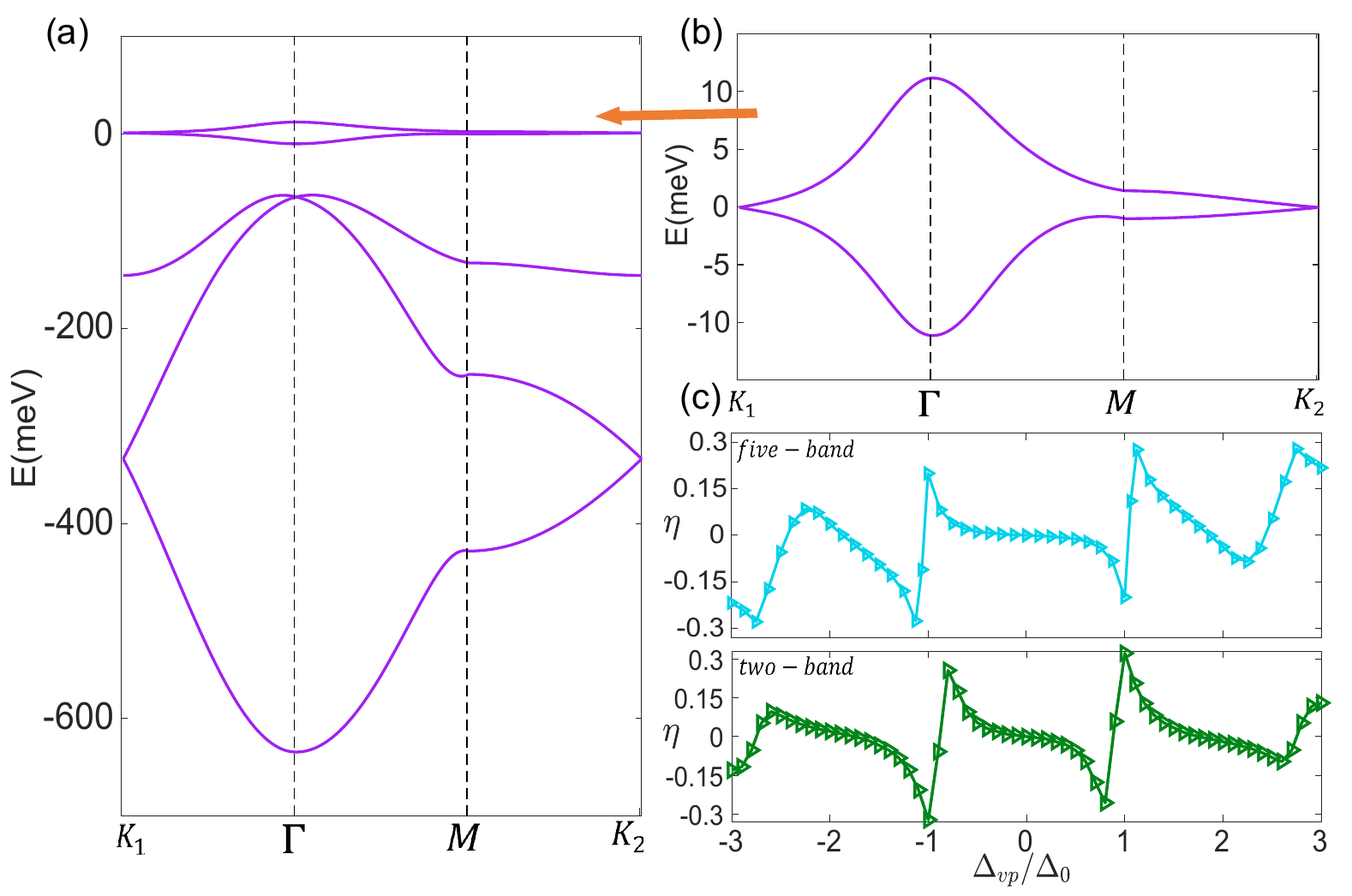}
	\caption{(a) The calculated band structure from the five-band model. (b) A zoom-in for the two nearly flat bands. (c) The diode efficiency $\eta$ as a function of $\Delta_{vp}/\Delta_0$ calculated by the five-band model and two-band model. The parameters are the same as in the main text.}
	\label{fig:figS4}
\end{figure}
To some extent, the minimal two-band tight binding model may not fully describe TBG because of the Wannier obstruction. There is a Wannier obstruction in TBG if \cite{zou2018band}:

(1) Keeping only the two active bands per spin per valley

(2) There is $C_2 T$ symmetry.

(3) There is an emergent valley $U_v(1)$ symmetry.

To overcome the Wannier obstruction, we adopt the faithful five-band tight binding model developed by Po et. al. \cite{po2019faithful}. The five band model includes five orbitals per unit cell (three $p$-orbitals $p_z,p_{\pm}=p_x\pm ip_y$ on AA spot and two $s$-orbitals on AB, BA spot). The quasiorbital wavefunctions are denoted by $\hat{\rho}_{\bm{k}}^{(l)}$, where $l=1,2,3$ labels the three kagome sites in each unit cell. The $\hat{\rho}_{\bm{k}}^{(l)}$ reads:
\begin{equation}
	\hat{\rho}_{\bm{k}}^{(1)}=\left(\begin{array}{c}
		i \tilde{a} (\phi_{11}-\phi_{10}) \\
		\tilde{b}\phi_{11}+\tilde{c}\phi_{10}\\
		\tilde{c}\phi_{11}+\tilde{b}\phi_{10}\\
		\tilde{d}^* \phi_{10}\\
		\tilde{d}
	\end{array}\right), \  \hat{\rho}_{\bm{k}}^{(2)}=\left(\begin{array}{c}
		i \tilde{a} (1-\phi_{11}) \\
		\omega(\tilde{b}+\tilde{c}\phi_{11})\\
		\omega^*(\tilde{c}+\tilde{b}\phi_{11})\\
		\tilde{d}^* \\
		\tilde{d}
	\end{array}\right), \   \hat{\rho}_{\bm{k}}^{(3)}=\left(\begin{array}{c}
		i \tilde{a} (\phi_{10}-1) \\
		\omega^*(\tilde{b}\phi_{10}+\tilde{c})\\
		\omega(\tilde{c}\phi_{10}+\tilde{b})\\
		\tilde{d}^*\phi_{0\bar{1}} \\
		\tilde{d}
	\end{array}\right),
\end{equation}
where $\tilde{a},\tilde{b},\tilde{c}$ are real and $\tilde{d}$ can be complex. $\omega=e^{i2\pi/3}$. $\phi_{lm}=e^{-i\bm{k}\cdot (l\bm{a_1}+m\bm{a_2})}$ and the negative numbers are denoted by $\bar{l}=-l$. The lattice vectors are $\bm{a_1}=(1/2,-\sqrt{3}/2)L_M,\bm{a_2}=(0,1)L_M$. After aggregating the three column vectors into a $5\times 3$ matrix $\rho_{\bm{k}}=(\rho_{\bm{k}}^{(1)},\rho_{\bm{k}}^{(2)},\rho_{\bm{k}}^{(3)})$, the five-band tight binding Hamiltonian in the Bloch basis can be written as
\begin{equation}
	H=-t_0 \rho_{\bm{k}}\rho_{\bm{k}}^{\dagger}+\text{diag}(\mu_{p_z},\mu_{p_{\pm}},\mu_{p_{\pm}},\mu_{s},\mu_{s}).
\end{equation}
In our calculation, the parameters are set as $(\tilde{a},\tilde{b},\tilde{c},\tilde{d})=(0.25,0.2,0.1,0.67)$, $(\mu_{p_z},\mu_{p_{\pm}},\mu_{s})=(-0.043,0,0.05)t_0$. $t_0=240$ meV. More details of the five-band model are in Ref.\cite{po2019faithful}.

The band structures from the five-band model are shown in Fig.\ref{fig:figS4}~(a),(b). We adopt the five-band model to study the Josephson diode effect in TBG following the same procedures as the two-band model, where we set the pairing to be the intra-orbital onsite $s$-wave. The warping term is already built-in the five-band model and there is no fine-tuning of parameters. The parameters of the five-band model are chosen in such a way that the bandwidths of the flat bands at the Fermi energy match the bandwidths of the flat bands in the two-band model.  The diode efficiency as a function of valley polarization Fig.\ref{fig:figS4}~(c) is similar to the calculation by the minimal two-band model  quantitatively.

\section{IV. Josephson current in the valley-polarized Chern insulator phase of TBG}
\begin{figure}[h]
  \centering
  \includegraphics[width=0.7
\linewidth]{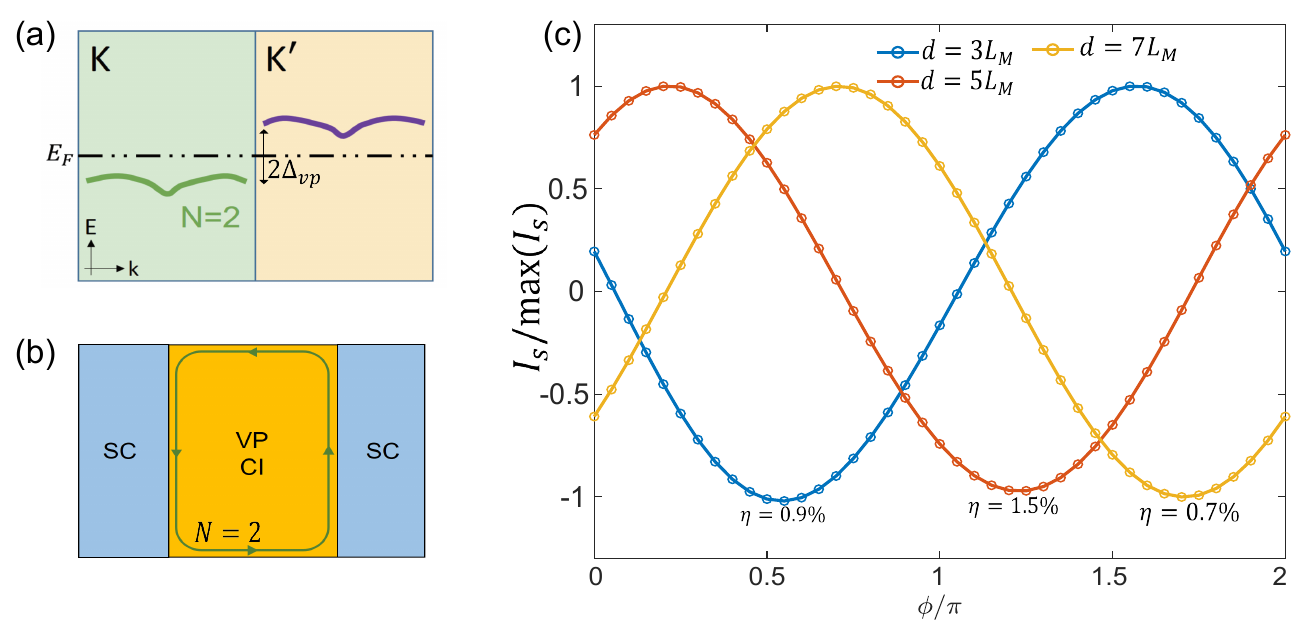}
  \caption{(a) Schematic picture of the valley-polarized Chern insulating state in TBG (\addYM{each band is  spin degenerated}), $E_F$ is the Fermi energy, $K$ and $K'$ label the opposite valley. (b) The Josephson current is mainly mediated by the chiral edge states around the boundary of the weak link. (c) The supercurrent density  (normalized by its maximal value) versus the phase difference $\phi$ with junction length $d$, where the junction region is set to be the valley-polarized Chern insulating states with $N=2$ at half-filling with $\Delta_{vp}=10\Delta_0$. The temperature is set to $k_B T=0.15\Delta_0$.}
  \label{fig:figS5}
\end{figure}
When the valley polarization $\Delta_{vp} \gg \Delta_0 $, in the case that the weak link is at half-filling and only one valley is occupied, the weak link region could be a Chern insulator with Chern number $N=2$ \cite{das2021symmetry,lian2021twisted,diez2021magnetic}. The $N=2$ is due to the spin degrees of freedom in each valley. In this case, the Josephson current is mainly mediated by the chiral edge states around the boundary of the weak link [Fig.\ref{fig:figS5}(a)] as the bulk is fully gapped. To model the Chern insulator at the weak link, a Haldane term is added to our original tight binding Hamiltonian. The Haldane term will not affect the results in any significant way when the junction is metallic. When the bulk is insulating, edge states generated by the Haldane term will mediate the supercurrent across the junction. The Haldane term can be written as
\begin{equation}
H_h=-m_0 \sum_i (-1)^{X(i)}c^\dagger_{i;\tau\sigma}c_{j;\tau\sigma}-\sum_{\ll ij\gg} t_X^\tau c^\dagger_{i;\tau\sigma}c_{j;\tau\sigma}+h.c.
\end{equation}
Where $m_0$ is the mass term, $X(i)=\pm 1$ on the A and B sublattices and $t_A^\tau,t_B^\tau$ are the second-neighbor hopping terms. Because of the time reversal symmetry, $t_X^-=(t_X^+)^*$. Without loss of generality, we set $t_A^+=0.2t_1e^{i\pi/2}$, $t_B^+=0.2t_1e^{-i\pi/2}$ and $m_0=0.1t_1$.

The resulting current phase relation is shown in Fig.\ref{fig:figS5}(b). Three plots are shown for junctions with different width $d$ (the junction widths are 3, 5 and 7 moir\'{e} unit cells respectively). It is clear that the current phase relation is still unconventional in the sense that $I_s=\sin(\phi-\phi_0)$ with a finite $\phi_0$. In other words, the junctions are $\phi_0$-Josephson junctions. However, due to the dominant sinusoidal behavior, the diode effect is almost negligible. For the junctions with three different lengths shown in Fig.\ref{fig:figS5}~(c), the diode efficiencies range from 0.5\% to 2\% which are much smaller than the case when the weak-link is metallic. This result is also highly consistent with the experimental results that the diode effect is exceedingly weak when the edge states dominate the transport.
\section{V. Josephson diode effect in Rashba wire with anti-symmetric SOC}
\begin{figure}[h]
	\centering
	\includegraphics[width=0.6
	\linewidth]{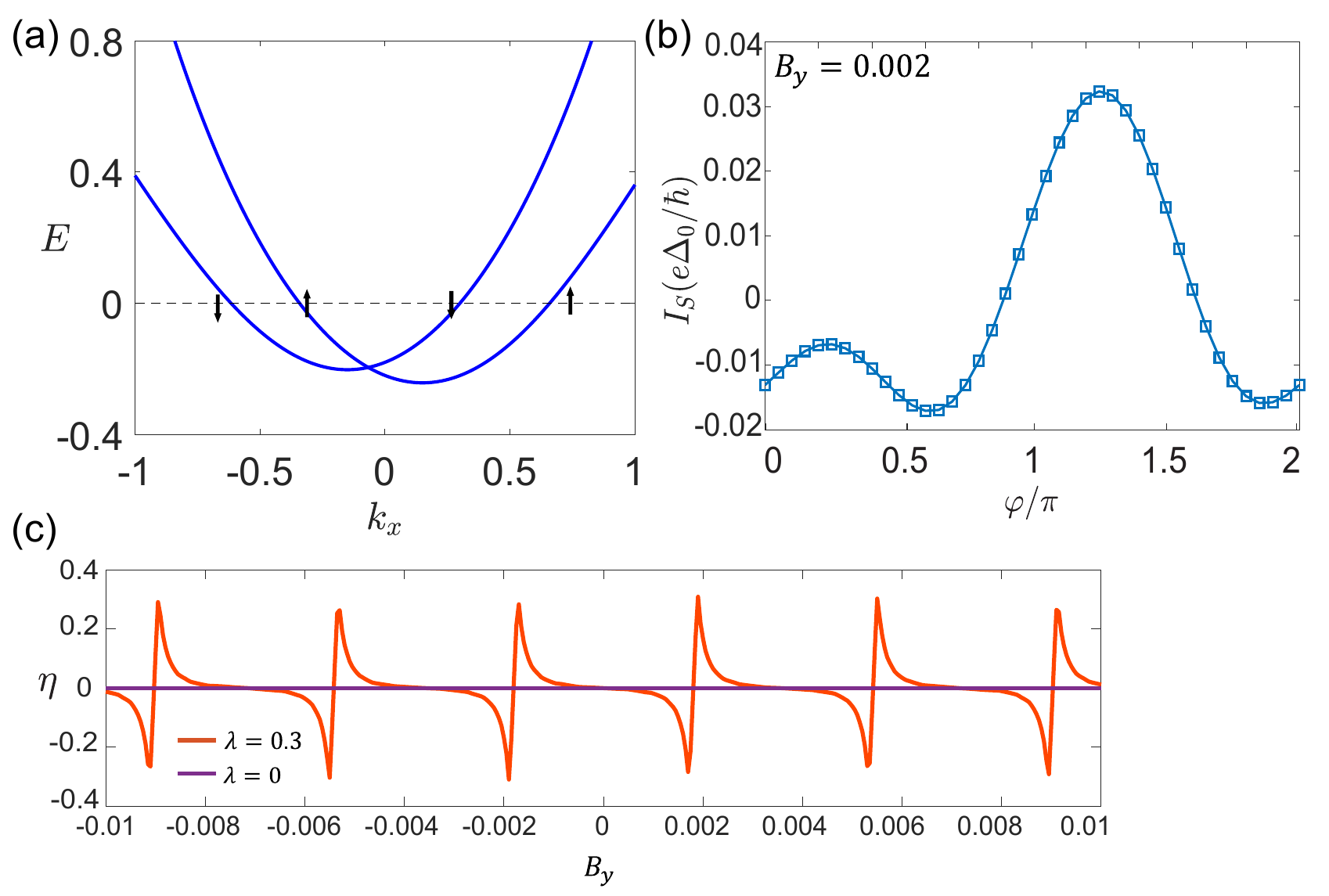}
	\caption{(a) The energy bands of a Rashba wire with $B_y=0.02$. The spin directions are labelled by black arrows. (b) The nonreciprocal current-phase relation at $B_y=0.002$. (c) The nonreciprocity efficiency $\eta$ as a function of $B_y$ with and without ASOC. The pairing potential $\Delta_0=0.004$. The distance of the weak-link region is $d=200$.}
	\label{fig:figS6}
\end{figure}
In this section we propose a 1D Rashba wire with anti-symmetric spin-orbit coupling (ASOC) to realize the Josephson diode effect when an in-plane magnetic field is applied in the weak-link region. The model Hamiltonian can be written as
\begin{equation}
h=(t k_x^2-\mu)\sigma_0+\alpha_R k_x\sigma_y +\lambda k_x^3\sigma_z + B_y\sigma_y,
\end{equation}
where $\alpha_R$ denotes the strength of Rashba SOC, and $\lambda$ denotes the strength of ASOC\cite{fu2009hexagonal}. In our calculations we set $t=1$, $\alpha_R=0.3$, $\lambda=0.3$. The energy bands are shown in Fig.\ref{fig:figS6}~(a) with finite $B_y$. In Fig.\ref{fig:figS6}~(b) we calculate the current-phase relation with $B_y=0.002$, which displays a large nonreciprocity efficiency $\eta\sim 30\%$. In Fig.\ref{fig:figS6}~(c) we show the $\eta$ as a function of $B_y$ with and without $\lambda$ term. It is clear that the in-plane magnetic field can be mapped to the valley polarization and the ASOC can be mapped to the trigonal warping effect. Thus our theory is not limited in valley-polarized system and can also be applied to the materials with strong SOC.


\begin{thebibliography}{51}%
\makeatletter
\providecommand \@ifxundefined [1]{%
 \@ifx{#1\undefined}
}%
\providecommand \@ifnum [1]{%
 \ifnum #1\expandafter \@firstoftwo
 \else \expandafter \@secondoftwo
 \fi
}%
\providecommand \@ifx [1]{%
 \ifx #1\expandafter \@firstoftwo
 \else \expandafter \@secondoftwo
 \fi
}%
\providecommand \natexlab [1]{#1}%
\providecommand \enquote  [1]{``#1''}%
\providecommand \bibnamefont  [1]{#1}%
\providecommand \bibfnamefont [1]{#1}%
\providecommand \citenamefont [1]{#1}%
\providecommand \href@noop [0]{\@secondoftwo}%
\providecommand \href [0]{\begingroup \@sanitize@url \@href}%
\providecommand \@href[1]{\@@startlink{#1}\@@href}%
\providecommand \@@href[1]{\endgroup#1\@@endlink}%
\providecommand \@sanitize@url [0]{\catcode `\\12\catcode `\$12\catcode
  `\&12\catcode `\#12\catcode `\^12\catcode `\_12\catcode `\%12\relax}%
\providecommand \@@startlink[1]{}%
\providecommand \@@endlink[0]{}%
\providecommand \url  [0]{\begingroup\@sanitize@url \@url }%
\providecommand \@url [1]{\endgroup\@href {#1}{\urlprefix }}%
\providecommand \urlprefix  [0]{URL }%
\providecommand \Eprint [0]{\href }%
\providecommand \doibase [0]{http://dx.doi.org/}%
\providecommand \selectlanguage [0]{\@gobble}%
\providecommand \bibinfo  [0]{\@secondoftwo}%
\providecommand \bibfield  [0]{\@secondoftwo}%
\providecommand \translation [1]{[#1]}%
\providecommand \BibitemOpen [0]{}%
\providecommand \bibitemStop [0]{}%
\providecommand \bibitemNoStop [0]{.\EOS\space}%
\providecommand \EOS [0]{\spacefactor3000\relax}%
\providecommand \BibitemShut  [1]{\csname bibitem#1\endcsname}%
\let\auto@bib@innerbib\@empty
\bibitem [{\citenamefont {Josephson}(1962)}]{josephson1962possible}%
  \BibitemOpen
  \bibfield  {author} {\bibinfo {author} {\bibfnamefont {B.~D.}\ \bibnamefont
  {Josephson}},\ }\href@noop {} {\bibfield  {journal} {\bibinfo  {journal}
  {Physics letters}\ }\textbf {\bibinfo {volume} {1}},\ \bibinfo {pages} {251}
  (\bibinfo {year} {1962})}\BibitemShut {NoStop}%
\bibitem [{\citenamefont {Josephson}(1964)}]{josephson1964coupled}%
  \BibitemOpen
  \bibfield  {author} {\bibinfo {author} {\bibfnamefont {B.}~\bibnamefont
  {Josephson}},\ }\href@noop {} {\bibfield  {journal} {\bibinfo  {journal}
  {Reviews of Modern Physics}\ }\textbf {\bibinfo {volume} {36}},\ \bibinfo
  {pages} {216} (\bibinfo {year} {1964})}\BibitemShut {NoStop}%
\bibitem [{\citenamefont {Anderson}\ and\ \citenamefont
  {Rowell}(1963)}]{anderson1963probable}%
  \BibitemOpen
  \bibfield  {author} {\bibinfo {author} {\bibfnamefont {P.~W.}\ \bibnamefont
  {Anderson}}\ and\ \bibinfo {author} {\bibfnamefont {J.~M.}\ \bibnamefont
  {Rowell}},\ }\href@noop {} {\bibfield  {journal} {\bibinfo  {journal}
  {Physical Review Letters}\ }\textbf {\bibinfo {volume} {10}},\ \bibinfo
  {pages} {230} (\bibinfo {year} {1963})}\BibitemShut {NoStop}%
\bibitem [{\citenamefont {Ambegaokar}\ and\ \citenamefont
  {Baratoff}(1963)}]{ambegaokar1963tunneling}%
  \BibitemOpen
  \bibfield  {author} {\bibinfo {author} {\bibfnamefont {V.}~\bibnamefont
  {Ambegaokar}}\ and\ \bibinfo {author} {\bibfnamefont {A.}~\bibnamefont
  {Baratoff}},\ }\href@noop {} {\bibfield  {journal} {\bibinfo  {journal}
  {Physical Review Letters}\ }\textbf {\bibinfo {volume} {10}},\ \bibinfo
  {pages} {486} (\bibinfo {year} {1963})}\BibitemShut {NoStop}%
\bibitem [{\citenamefont {Feofanov}\ \emph {et~al.}(2010)\citenamefont
  {Feofanov}, \citenamefont {Oboznov}, \citenamefont {Bol’Ginov},
  \citenamefont {Lisenfeld}, \citenamefont {Poletto}, \citenamefont {Ryazanov},
  \citenamefont {Rossolenko}, \citenamefont {Khabipov}, \citenamefont
  {Balashov}, \citenamefont {Zorin} \emph
  {et~al.}}]{feofanov2010implementation}%
  \BibitemOpen
  \bibfield  {author} {\bibinfo {author} {\bibfnamefont {A.}~\bibnamefont
  {Feofanov}}, \bibinfo {author} {\bibfnamefont {V.}~\bibnamefont {Oboznov}},
  \bibinfo {author} {\bibfnamefont {V.}~\bibnamefont {Bol’Ginov}}, \bibinfo
  {author} {\bibfnamefont {J.}~\bibnamefont {Lisenfeld}}, \bibinfo {author}
  {\bibfnamefont {S.}~\bibnamefont {Poletto}}, \bibinfo {author} {\bibfnamefont
  {V.}~\bibnamefont {Ryazanov}}, \bibinfo {author} {\bibfnamefont
  {A.}~\bibnamefont {Rossolenko}}, \bibinfo {author} {\bibfnamefont
  {M.}~\bibnamefont {Khabipov}}, \bibinfo {author} {\bibfnamefont
  {D.}~\bibnamefont {Balashov}}, \bibinfo {author} {\bibfnamefont
  {A.}~\bibnamefont {Zorin}},  \emph {et~al.},\ }\href@noop {} {\bibfield
  {journal} {\bibinfo  {journal} {Nature Physics}\ }\textbf {\bibinfo {volume}
  {6}},\ \bibinfo {pages} {593} (\bibinfo {year} {2010})}\BibitemShut {NoStop}%
\bibitem [{\citenamefont {Yamashita}\ \emph {et~al.}(2005)\citenamefont
  {Yamashita}, \citenamefont {Tanikawa}, \citenamefont {Takahashi},\ and\
  \citenamefont {Maekawa}}]{yamashita2005superconducting}%
  \BibitemOpen
  \bibfield  {author} {\bibinfo {author} {\bibfnamefont {T.}~\bibnamefont
  {Yamashita}}, \bibinfo {author} {\bibfnamefont {K.}~\bibnamefont {Tanikawa}},
  \bibinfo {author} {\bibfnamefont {S.}~\bibnamefont {Takahashi}}, \ and\
  \bibinfo {author} {\bibfnamefont {S.}~\bibnamefont {Maekawa}},\ }\href@noop
  {} {\bibfield  {journal} {\bibinfo  {journal} {Physical review letters}\
  }\textbf {\bibinfo {volume} {95}},\ \bibinfo {pages} {097001} (\bibinfo
  {year} {2005})}\BibitemShut {NoStop}%
\bibitem [{\citenamefont {Kato}\ \emph {et~al.}(2007)\citenamefont {Kato},
  \citenamefont {Golubov},\ and\ \citenamefont
  {Nakamura}}]{kato2007decoherence}%
  \BibitemOpen
  \bibfield  {author} {\bibinfo {author} {\bibfnamefont {T.}~\bibnamefont
  {Kato}}, \bibinfo {author} {\bibfnamefont {A.~A.}\ \bibnamefont {Golubov}}, \
  and\ \bibinfo {author} {\bibfnamefont {Y.}~\bibnamefont {Nakamura}},\
  }\href@noop {} {\bibfield  {journal} {\bibinfo  {journal} {Physical Review
  B}\ }\textbf {\bibinfo {volume} {76}},\ \bibinfo {pages} {172502} (\bibinfo
  {year} {2007})}\BibitemShut {NoStop}%
\bibitem [{\citenamefont {Yamashita}\ \emph {et~al.}(2006)\citenamefont
  {Yamashita}, \citenamefont {Takahashi},\ and\ \citenamefont
  {Maekawa}}]{yamashita2006superconducting}%
  \BibitemOpen
  \bibfield  {author} {\bibinfo {author} {\bibfnamefont {T.}~\bibnamefont
  {Yamashita}}, \bibinfo {author} {\bibfnamefont {S.}~\bibnamefont
  {Takahashi}}, \ and\ \bibinfo {author} {\bibfnamefont {S.}~\bibnamefont
  {Maekawa}},\ }\href@noop {} {\bibfield  {journal} {\bibinfo  {journal}
  {Applied physics letters}\ }\textbf {\bibinfo {volume} {88}},\ \bibinfo
  {pages} {132501} (\bibinfo {year} {2006})}\BibitemShut {NoStop}%
\bibitem [{\citenamefont {Dolcini}\ \emph {et~al.}(2015)\citenamefont
  {Dolcini}, \citenamefont {Houzet},\ and\ \citenamefont
  {Meyer}}]{dolcini2015topological}%
  \BibitemOpen
  \bibfield  {author} {\bibinfo {author} {\bibfnamefont {F.}~\bibnamefont
  {Dolcini}}, \bibinfo {author} {\bibfnamefont {M.}~\bibnamefont {Houzet}}, \
  and\ \bibinfo {author} {\bibfnamefont {J.~S.}\ \bibnamefont {Meyer}},\
  }\href@noop {} {\bibfield  {journal} {\bibinfo  {journal} {Physical Review
  B}\ }\textbf {\bibinfo {volume} {92}},\ \bibinfo {pages} {035428} (\bibinfo
  {year} {2015})}\BibitemShut {NoStop}%
\bibitem [{\citenamefont {Chen}\ \emph {et~al.}(2018)\citenamefont {Chen},
  \citenamefont {He}, \citenamefont {Ali}, \citenamefont {Lee}, \citenamefont
  {Fong},\ and\ \citenamefont {Law}}]{chen2018asymmetric}%
  \BibitemOpen
  \bibfield  {author} {\bibinfo {author} {\bibfnamefont {C.-Z.}\ \bibnamefont
  {Chen}}, \bibinfo {author} {\bibfnamefont {J.~J.}\ \bibnamefont {He}},
  \bibinfo {author} {\bibfnamefont {M.~N.}\ \bibnamefont {Ali}}, \bibinfo
  {author} {\bibfnamefont {G.-H.}\ \bibnamefont {Lee}}, \bibinfo {author}
  {\bibfnamefont {K.~C.}\ \bibnamefont {Fong}}, \ and\ \bibinfo {author}
  {\bibfnamefont {K.~T.}\ \bibnamefont {Law}},\ }\href@noop {} {\bibfield
  {journal} {\bibinfo  {journal} {Physical Review B}\ }\textbf {\bibinfo
  {volume} {98}},\ \bibinfo {pages} {075430} (\bibinfo {year}
  {2018})}\BibitemShut {NoStop}%
\bibitem [{\citenamefont {Davydova}\ \emph {et~al.}(2022)\citenamefont
  {Davydova}, \citenamefont {Prembabu},\ and\ \citenamefont
  {Fu}}]{davydova2022universal}%
  \BibitemOpen
  \bibfield  {author} {\bibinfo {author} {\bibfnamefont {M.}~\bibnamefont
  {Davydova}}, \bibinfo {author} {\bibfnamefont {S.}~\bibnamefont {Prembabu}},
  \ and\ \bibinfo {author} {\bibfnamefont {L.}~\bibnamefont {Fu}},\ }\href@noop
  {} {\bibfield  {journal} {\bibinfo  {journal} {Science advances}\ }\textbf
  {\bibinfo {volume} {8}},\ \bibinfo {pages} {eabo0309} (\bibinfo {year}
  {2022})}\BibitemShut {NoStop}%
\bibitem [{\citenamefont {Zhang}\ \emph
  {et~al.}(2022{\natexlab{a}})\citenamefont {Zhang}, \citenamefont {Gu},
  \citenamefont {Li}, \citenamefont {Hu},\ and\ \citenamefont
  {Jiang}}]{zhang2022general}%
  \BibitemOpen
  \bibfield  {author} {\bibinfo {author} {\bibfnamefont {Y.}~\bibnamefont
  {Zhang}}, \bibinfo {author} {\bibfnamefont {Y.}~\bibnamefont {Gu}}, \bibinfo
  {author} {\bibfnamefont {P.}~\bibnamefont {Li}}, \bibinfo {author}
  {\bibfnamefont {J.}~\bibnamefont {Hu}}, \ and\ \bibinfo {author}
  {\bibfnamefont {K.}~\bibnamefont {Jiang}},\ }\href@noop {} {\bibfield
  {journal} {\bibinfo  {journal} {Physical Review X}\ }\textbf {\bibinfo
  {volume} {12}},\ \bibinfo {pages} {041013} (\bibinfo {year}
  {2022}{\natexlab{a}})}\BibitemShut {NoStop}%
\bibitem [{\citenamefont {Tanaka}\ \emph {et~al.}(2022)\citenamefont {Tanaka},
  \citenamefont {Lu},\ and\ \citenamefont {Nagaosa}}]{tanaka2022theory}%
  \BibitemOpen
  \bibfield  {author} {\bibinfo {author} {\bibfnamefont {Y.}~\bibnamefont
  {Tanaka}}, \bibinfo {author} {\bibfnamefont {B.}~\bibnamefont {Lu}}, \ and\
  \bibinfo {author} {\bibfnamefont {N.}~\bibnamefont {Nagaosa}},\ }\href@noop
  {} {\bibfield  {journal} {\bibinfo  {journal} {Physical Review B}\ }\textbf
  {\bibinfo {volume} {106}},\ \bibinfo {pages} {214524} (\bibinfo {year}
  {2022})}\BibitemShut {NoStop}%
\bibitem [{\citenamefont {Lu}\ \emph {et~al.}(2022)\citenamefont {Lu},
  \citenamefont {Ikegaya}, \citenamefont {Burset}, \citenamefont {Tanaka},\
  and\ \citenamefont {Nagaosa}}]{lu2022josephson}%
  \BibitemOpen
  \bibfield  {author} {\bibinfo {author} {\bibfnamefont {B.}~\bibnamefont
  {Lu}}, \bibinfo {author} {\bibfnamefont {S.}~\bibnamefont {Ikegaya}},
  \bibinfo {author} {\bibfnamefont {P.}~\bibnamefont {Burset}}, \bibinfo
  {author} {\bibfnamefont {Y.}~\bibnamefont {Tanaka}}, \ and\ \bibinfo {author}
  {\bibfnamefont {N.}~\bibnamefont {Nagaosa}},\ }\href@noop {} {\bibfield
  {journal} {\bibinfo  {journal} {arXiv preprint arXiv:2211.10572}\ } (\bibinfo
  {year} {2022})}\BibitemShut {NoStop}%
\bibitem [{\citenamefont {Wang}\ \emph {et~al.}(2022)\citenamefont {Wang},
  \citenamefont {Wang},\ and\ \citenamefont {Wu}}]{wang2022symmetry}%
  \BibitemOpen
  \bibfield  {author} {\bibinfo {author} {\bibfnamefont {D.}~\bibnamefont
  {Wang}}, \bibinfo {author} {\bibfnamefont {Q.-H.}\ \bibnamefont {Wang}}, \
  and\ \bibinfo {author} {\bibfnamefont {C.}~\bibnamefont {Wu}},\ }\href@noop
  {} {\bibfield  {journal} {\bibinfo  {journal} {arXiv preprint
  arXiv:2209.12646}\ } (\bibinfo {year} {2022})}\BibitemShut {NoStop}%
\bibitem [{\citenamefont {Ando}\ \emph {et~al.}(2020)\citenamefont {Ando},
  \citenamefont {Miyasaka}, \citenamefont {Li}, \citenamefont {Ishizuka},
  \citenamefont {Arakawa}, \citenamefont {Shiota}, \citenamefont {Moriyama},
  \citenamefont {Yanase},\ and\ \citenamefont {Ono}}]{ando2020observation}%
  \BibitemOpen
  \bibfield  {author} {\bibinfo {author} {\bibfnamefont {F.}~\bibnamefont
  {Ando}}, \bibinfo {author} {\bibfnamefont {Y.}~\bibnamefont {Miyasaka}},
  \bibinfo {author} {\bibfnamefont {T.}~\bibnamefont {Li}}, \bibinfo {author}
  {\bibfnamefont {J.}~\bibnamefont {Ishizuka}}, \bibinfo {author}
  {\bibfnamefont {T.}~\bibnamefont {Arakawa}}, \bibinfo {author} {\bibfnamefont
  {Y.}~\bibnamefont {Shiota}}, \bibinfo {author} {\bibfnamefont
  {T.}~\bibnamefont {Moriyama}}, \bibinfo {author} {\bibfnamefont
  {Y.}~\bibnamefont {Yanase}}, \ and\ \bibinfo {author} {\bibfnamefont
  {T.}~\bibnamefont {Ono}},\ }\href@noop {} {\bibfield  {journal} {\bibinfo
  {journal} {Nature}\ }\textbf {\bibinfo {volume} {584}},\ \bibinfo {pages}
  {373} (\bibinfo {year} {2020})}\BibitemShut {NoStop}%
\bibitem [{\citenamefont {Misaki}\ and\ \citenamefont
  {Nagaosa}(2021)}]{misaki2021theory}%
  \BibitemOpen
  \bibfield  {author} {\bibinfo {author} {\bibfnamefont {K.}~\bibnamefont
  {Misaki}}\ and\ \bibinfo {author} {\bibfnamefont {N.}~\bibnamefont
  {Nagaosa}},\ }\href@noop {} {\bibfield  {journal} {\bibinfo  {journal}
  {Physical Review B}\ }\textbf {\bibinfo {volume} {103}},\ \bibinfo {pages}
  {245302} (\bibinfo {year} {2021})}\BibitemShut {NoStop}%
\bibitem [{\citenamefont {Rymarz}\ \emph {et~al.}(2021)\citenamefont {Rymarz},
  \citenamefont {Bosco}, \citenamefont {Ciani},\ and\ \citenamefont
  {DiVincenzo}}]{rymarz2021hardware}%
  \BibitemOpen
  \bibfield  {author} {\bibinfo {author} {\bibfnamefont {M.}~\bibnamefont
  {Rymarz}}, \bibinfo {author} {\bibfnamefont {S.}~\bibnamefont {Bosco}},
  \bibinfo {author} {\bibfnamefont {A.}~\bibnamefont {Ciani}}, \ and\ \bibinfo
  {author} {\bibfnamefont {D.~P.}\ \bibnamefont {DiVincenzo}},\ }\href@noop {}
  {\bibfield  {journal} {\bibinfo  {journal} {Physical Review X}\ }\textbf
  {\bibinfo {volume} {11}},\ \bibinfo {pages} {011032} (\bibinfo {year}
  {2021})}\BibitemShut {NoStop}%
\bibitem [{\citenamefont {Wu}\ \emph {et~al.}(2022)\citenamefont {Wu},
  \citenamefont {Wang}, \citenamefont {Xu}, \citenamefont {Sivakumar},
  \citenamefont {Pasco}, \citenamefont {Filippozzi}, \citenamefont {Parkin},
  \citenamefont {Zeng}, \citenamefont {McQueen},\ and\ \citenamefont
  {Ali}}]{wu2022field}%
  \BibitemOpen
  \bibfield  {author} {\bibinfo {author} {\bibfnamefont {H.}~\bibnamefont
  {Wu}}, \bibinfo {author} {\bibfnamefont {Y.}~\bibnamefont {Wang}}, \bibinfo
  {author} {\bibfnamefont {Y.}~\bibnamefont {Xu}}, \bibinfo {author}
  {\bibfnamefont {P.~K.}\ \bibnamefont {Sivakumar}}, \bibinfo {author}
  {\bibfnamefont {C.}~\bibnamefont {Pasco}}, \bibinfo {author} {\bibfnamefont
  {U.}~\bibnamefont {Filippozzi}}, \bibinfo {author} {\bibfnamefont {S.~S.}\
  \bibnamefont {Parkin}}, \bibinfo {author} {\bibfnamefont {Y.-J.}\
  \bibnamefont {Zeng}}, \bibinfo {author} {\bibfnamefont {T.}~\bibnamefont
  {McQueen}}, \ and\ \bibinfo {author} {\bibfnamefont {M.~N.}\ \bibnamefont
  {Ali}},\ }\href@noop {} {\bibfield  {journal} {\bibinfo  {journal} {Nature}\
  }\textbf {\bibinfo {volume} {604}},\ \bibinfo {pages} {653} (\bibinfo {year}
  {2022})}\BibitemShut {NoStop}%
\bibitem [{\citenamefont {Pal}\ \emph {et~al.}(2022)\citenamefont {Pal},
  \citenamefont {Chakraborty}, \citenamefont {Sivakumar}, \citenamefont
  {Davydova}, \citenamefont {Gopi}, \citenamefont {Pandeya}, \citenamefont
  {Krieger}, \citenamefont {Zhang}, \citenamefont {Date}, \citenamefont {Ju}
  \emph {et~al.}}]{pal2022josephson}%
  \BibitemOpen
  \bibfield  {author} {\bibinfo {author} {\bibfnamefont {B.}~\bibnamefont
  {Pal}}, \bibinfo {author} {\bibfnamefont {A.}~\bibnamefont {Chakraborty}},
  \bibinfo {author} {\bibfnamefont {P.~K.}\ \bibnamefont {Sivakumar}}, \bibinfo
  {author} {\bibfnamefont {M.}~\bibnamefont {Davydova}}, \bibinfo {author}
  {\bibfnamefont {A.~K.}\ \bibnamefont {Gopi}}, \bibinfo {author}
  {\bibfnamefont {A.~K.}\ \bibnamefont {Pandeya}}, \bibinfo {author}
  {\bibfnamefont {J.~A.}\ \bibnamefont {Krieger}}, \bibinfo {author}
  {\bibfnamefont {Y.}~\bibnamefont {Zhang}}, \bibinfo {author} {\bibfnamefont
  {M.}~\bibnamefont {Date}}, \bibinfo {author} {\bibfnamefont {S.}~\bibnamefont
  {Ju}},  \emph {et~al.},\ }\href@noop {} {\bibfield  {journal} {\bibinfo
  {journal} {Nature physics}\ }\textbf {\bibinfo {volume} {18}},\ \bibinfo
  {pages} {1228} (\bibinfo {year} {2022})}\BibitemShut {NoStop}%
\bibitem [{\citenamefont {Diez-Merida}\ \emph {et~al.}(2021)\citenamefont
  {Diez-Merida}, \citenamefont {D{\'\i}ez-Carl{\'o}n}, \citenamefont {Yang},
  \citenamefont {Xie}, \citenamefont {Gao}, \citenamefont {Watanabe},
  \citenamefont {Taniguchi}, \citenamefont {Lu}, \citenamefont {Law},\ and\
  \citenamefont {Efetov}}]{diez2021magnetic}%
  \BibitemOpen
  \bibfield  {author} {\bibinfo {author} {\bibfnamefont {J.}~\bibnamefont
  {Diez-Merida}}, \bibinfo {author} {\bibfnamefont {A.}~\bibnamefont
  {D{\'\i}ez-Carl{\'o}n}}, \bibinfo {author} {\bibfnamefont {S.}~\bibnamefont
  {Yang}}, \bibinfo {author} {\bibfnamefont {Y.-M.}\ \bibnamefont {Xie}},
  \bibinfo {author} {\bibfnamefont {X.-J.}\ \bibnamefont {Gao}}, \bibinfo
  {author} {\bibfnamefont {K.}~\bibnamefont {Watanabe}}, \bibinfo {author}
  {\bibfnamefont {T.}~\bibnamefont {Taniguchi}}, \bibinfo {author}
  {\bibfnamefont {X.}~\bibnamefont {Lu}}, \bibinfo {author} {\bibfnamefont
  {K.~T.}\ \bibnamefont {Law}}, \ and\ \bibinfo {author} {\bibfnamefont
  {D.~K.}\ \bibnamefont {Efetov}},\ }\href@noop {} {\bibfield  {journal}
  {\bibinfo  {journal} {arXiv preprint arXiv:2110.01067}\ } (\bibinfo {year}
  {2021})}\BibitemShut {NoStop}%
\bibitem [{\citenamefont {Baumgartner}\ \emph {et~al.}(2022)\citenamefont
  {Baumgartner}, \citenamefont {Fuchs}, \citenamefont {Costa}, \citenamefont
  {Reinhardt}, \citenamefont {Gronin}, \citenamefont {Gardner}, \citenamefont
  {Lindemann}, \citenamefont {Manfra}, \citenamefont {Faria~Junior},
  \citenamefont {Kochan} \emph {et~al.}}]{baumgartner2022supercurrent}%
  \BibitemOpen
  \bibfield  {author} {\bibinfo {author} {\bibfnamefont {C.}~\bibnamefont
  {Baumgartner}}, \bibinfo {author} {\bibfnamefont {L.}~\bibnamefont {Fuchs}},
  \bibinfo {author} {\bibfnamefont {A.}~\bibnamefont {Costa}}, \bibinfo
  {author} {\bibfnamefont {S.}~\bibnamefont {Reinhardt}}, \bibinfo {author}
  {\bibfnamefont {S.}~\bibnamefont {Gronin}}, \bibinfo {author} {\bibfnamefont
  {G.~C.}\ \bibnamefont {Gardner}}, \bibinfo {author} {\bibfnamefont
  {T.}~\bibnamefont {Lindemann}}, \bibinfo {author} {\bibfnamefont {M.~J.}\
  \bibnamefont {Manfra}}, \bibinfo {author} {\bibfnamefont {P.~E.}\
  \bibnamefont {Faria~Junior}}, \bibinfo {author} {\bibfnamefont
  {D.}~\bibnamefont {Kochan}},  \emph {et~al.},\ }\href@noop {} {\bibfield
  {journal} {\bibinfo  {journal} {Nature Nanotechnology}\ }\textbf {\bibinfo
  {volume} {17}},\ \bibinfo {pages} {39} (\bibinfo {year} {2022})}\BibitemShut
  {NoStop}%
\bibitem [{\citenamefont {Jeon}\ \emph {et~al.}(2022)\citenamefont {Jeon},
  \citenamefont {Kim}, \citenamefont {Yoon}, \citenamefont {Jeon},
  \citenamefont {Han}, \citenamefont {Cottet}, \citenamefont {Kontos},\ and\
  \citenamefont {Parkin}}]{jeon2022zero}%
  \BibitemOpen
  \bibfield  {author} {\bibinfo {author} {\bibfnamefont {K.-R.}\ \bibnamefont
  {Jeon}}, \bibinfo {author} {\bibfnamefont {J.-K.}\ \bibnamefont {Kim}},
  \bibinfo {author} {\bibfnamefont {J.}~\bibnamefont {Yoon}}, \bibinfo {author}
  {\bibfnamefont {J.-C.}\ \bibnamefont {Jeon}}, \bibinfo {author}
  {\bibfnamefont {H.}~\bibnamefont {Han}}, \bibinfo {author} {\bibfnamefont
  {A.}~\bibnamefont {Cottet}}, \bibinfo {author} {\bibfnamefont
  {T.}~\bibnamefont {Kontos}}, \ and\ \bibinfo {author} {\bibfnamefont {S.~S.}\
  \bibnamefont {Parkin}},\ }\href@noop {} {\bibfield  {journal} {\bibinfo
  {journal} {Nature Materials}\ }\textbf {\bibinfo {volume} {21}},\ \bibinfo
  {pages} {1008} (\bibinfo {year} {2022})}\BibitemShut {NoStop}%
\bibitem [{\citenamefont {Yuan}\ and\ \citenamefont
  {Fu}(2018)}]{yuan2018model}%
  \BibitemOpen
  \bibfield  {author} {\bibinfo {author} {\bibfnamefont {N.~F.}\ \bibnamefont
  {Yuan}}\ and\ \bibinfo {author} {\bibfnamefont {L.}~\bibnamefont {Fu}},\
  }\href@noop {} {\bibfield  {journal} {\bibinfo  {journal} {Physical Review
  B}\ }\textbf {\bibinfo {volume} {98}},\ \bibinfo {pages} {045103} (\bibinfo
  {year} {2018})}\BibitemShut {NoStop}%
\bibitem [{\citenamefont {Xie}\ \emph {et~al.}(2022)\citenamefont {Xie},
  \citenamefont {Efetov},\ and\ \citenamefont {Law}}]{xie2022valley}%
  \BibitemOpen
  \bibfield  {author} {\bibinfo {author} {\bibfnamefont {Y.-M.}\ \bibnamefont
  {Xie}}, \bibinfo {author} {\bibfnamefont {D.~K.}\ \bibnamefont {Efetov}}, \
  and\ \bibinfo {author} {\bibfnamefont {K.}~\bibnamefont {Law}},\ }\href@noop
  {} {\bibfield  {journal} {\bibinfo  {journal} {arXiv preprint
  arXiv:2202.05663}\ } (\bibinfo {year} {2022})}\BibitemShut {NoStop}%
\bibitem [{Not()}]{NoteX}%
  \BibitemOpen
  \href@noop {} {}\bibinfo {note} {See Supplementary Material for: 1.
  Scattering matrix method for the 1D toy model; 2. JDE in TBG with
  unconventional pairing; 3. Tight binding models for TBG; 4. Josephson
  current in the valley-polarized Chern insulator phase of TBG. 5. JDE in
  Rashba wire with anti-symmetric spin-orbit coupling.}\BibitemShut {Stop}%
\bibitem [{\citenamefont {Brouwer}\ and\ \citenamefont
  {Beenakker}(1997)}]{brouwer1997anomalous}%
  \BibitemOpen
  \bibfield  {author} {\bibinfo {author} {\bibfnamefont {P.}~\bibnamefont
  {Brouwer}}\ and\ \bibinfo {author} {\bibfnamefont {C.}~\bibnamefont
  {Beenakker}},\ }\href@noop {} {\bibfield  {journal} {\bibinfo  {journal}
  {Chaos, Solitons \& Fractals}\ }\textbf {\bibinfo {volume} {8}},\ \bibinfo
  {pages} {1249} (\bibinfo {year} {1997})}\BibitemShut {NoStop}%
\bibitem [{\citenamefont {Po}\ \emph {et~al.}(2018)\citenamefont {Po},
  \citenamefont {Zou}, \citenamefont {Vishwanath},\ and\ \citenamefont
  {Senthil}}]{po2018origin}%
  \BibitemOpen
  \bibfield  {author} {\bibinfo {author} {\bibfnamefont {H.~C.}\ \bibnamefont
  {Po}}, \bibinfo {author} {\bibfnamefont {L.}~\bibnamefont {Zou}}, \bibinfo
  {author} {\bibfnamefont {A.}~\bibnamefont {Vishwanath}}, \ and\ \bibinfo
  {author} {\bibfnamefont {T.}~\bibnamefont {Senthil}},\ }\href@noop {}
  {\bibfield  {journal} {\bibinfo  {journal} {Physical Review X}\ }\textbf
  {\bibinfo {volume} {8}},\ \bibinfo {pages} {031089} (\bibinfo {year}
  {2018})}\BibitemShut {NoStop}%
\bibitem [{\citenamefont {Lee}\ \emph {et~al.}(2019)\citenamefont {Lee},
  \citenamefont {Khalaf}, \citenamefont {Liu}, \citenamefont {Liu},
  \citenamefont {Hao}, \citenamefont {Kim},\ and\ \citenamefont
  {Vishwanath}}]{lee2019theory}%
  \BibitemOpen
  \bibfield  {author} {\bibinfo {author} {\bibfnamefont {J.~Y.}\ \bibnamefont
  {Lee}}, \bibinfo {author} {\bibfnamefont {E.}~\bibnamefont {Khalaf}},
  \bibinfo {author} {\bibfnamefont {S.}~\bibnamefont {Liu}}, \bibinfo {author}
  {\bibfnamefont {X.}~\bibnamefont {Liu}}, \bibinfo {author} {\bibfnamefont
  {Z.}~\bibnamefont {Hao}}, \bibinfo {author} {\bibfnamefont {P.}~\bibnamefont
  {Kim}}, \ and\ \bibinfo {author} {\bibfnamefont {A.}~\bibnamefont
  {Vishwanath}},\ }\href@noop {} {\bibfield  {journal} {\bibinfo  {journal}
  {Nature communications}\ }\textbf {\bibinfo {volume} {10}},\ \bibinfo {pages}
  {1} (\bibinfo {year} {2019})}\BibitemShut {NoStop}%
\bibitem [{\citenamefont {Bultinck}\ \emph {et~al.}(2020)\citenamefont
  {Bultinck}, \citenamefont {Chatterjee},\ and\ \citenamefont
  {Zaletel}}]{bultinck2020mechanism}%
  \BibitemOpen
  \bibfield  {author} {\bibinfo {author} {\bibfnamefont {N.}~\bibnamefont
  {Bultinck}}, \bibinfo {author} {\bibfnamefont {S.}~\bibnamefont
  {Chatterjee}}, \ and\ \bibinfo {author} {\bibfnamefont {M.~P.}\ \bibnamefont
  {Zaletel}},\ }\href@noop {} {\bibfield  {journal} {\bibinfo  {journal}
  {Physical review letters}\ }\textbf {\bibinfo {volume} {124}},\ \bibinfo
  {pages} {166601} (\bibinfo {year} {2020})}\BibitemShut {NoStop}%
\bibitem [{\citenamefont {Liu}\ and\ \citenamefont
  {Dai}(2021)}]{liu2021theories}%
  \BibitemOpen
  \bibfield  {author} {\bibinfo {author} {\bibfnamefont {J.}~\bibnamefont
  {Liu}}\ and\ \bibinfo {author} {\bibfnamefont {X.}~\bibnamefont {Dai}},\
  }\href@noop {} {\bibfield  {journal} {\bibinfo  {journal} {Physical Review
  B}\ }\textbf {\bibinfo {volume} {103}},\ \bibinfo {pages} {035427} (\bibinfo
  {year} {2021})}\BibitemShut {NoStop}%
\bibitem [{\citenamefont {Koshino}\ \emph {et~al.}(2018)\citenamefont
  {Koshino}, \citenamefont {Yuan}, \citenamefont {Koretsune}, \citenamefont
  {Ochi}, \citenamefont {Kuroki},\ and\ \citenamefont
  {Fu}}]{koshino2018maximally}%
  \BibitemOpen
  \bibfield  {author} {\bibinfo {author} {\bibfnamefont {M.}~\bibnamefont
  {Koshino}}, \bibinfo {author} {\bibfnamefont {N.~F.}\ \bibnamefont {Yuan}},
  \bibinfo {author} {\bibfnamefont {T.}~\bibnamefont {Koretsune}}, \bibinfo
  {author} {\bibfnamefont {M.}~\bibnamefont {Ochi}}, \bibinfo {author}
  {\bibfnamefont {K.}~\bibnamefont {Kuroki}}, \ and\ \bibinfo {author}
  {\bibfnamefont {L.}~\bibnamefont {Fu}},\ }\href@noop {} {\bibfield  {journal}
  {\bibinfo  {journal} {Physical Review X}\ }\textbf {\bibinfo {volume} {8}},\
  \bibinfo {pages} {031087} (\bibinfo {year} {2018})}\BibitemShut {NoStop}%
\bibitem [{\citenamefont {Pathak}\ \emph {et~al.}(2022)\citenamefont {Pathak},
  \citenamefont {Rakib}, \citenamefont {Hou}, \citenamefont {Nevidomskyy},
  \citenamefont {Ertekin}, \citenamefont {Johnson},\ and\ \citenamefont
  {Wagner}}]{pathak2022accurate}%
  \BibitemOpen
  \bibfield  {author} {\bibinfo {author} {\bibfnamefont {S.}~\bibnamefont
  {Pathak}}, \bibinfo {author} {\bibfnamefont {T.}~\bibnamefont {Rakib}},
  \bibinfo {author} {\bibfnamefont {R.}~\bibnamefont {Hou}}, \bibinfo {author}
  {\bibfnamefont {A.}~\bibnamefont {Nevidomskyy}}, \bibinfo {author}
  {\bibfnamefont {E.}~\bibnamefont {Ertekin}}, \bibinfo {author} {\bibfnamefont
  {H.~T.}\ \bibnamefont {Johnson}}, \ and\ \bibinfo {author} {\bibfnamefont
  {L.~K.}\ \bibnamefont {Wagner}},\ }\href@noop {} {\bibfield  {journal}
  {\bibinfo  {journal} {Physical Review B}\ }\textbf {\bibinfo {volume}
  {105}},\ \bibinfo {pages} {115141} (\bibinfo {year} {2022})}\BibitemShut
  {NoStop}%
\bibitem [{\citenamefont {Carr}\ \emph {et~al.}(2019)\citenamefont {Carr},
  \citenamefont {Fang}, \citenamefont {Zhu},\ and\ \citenamefont
  {Kaxiras}}]{carr2019exact}%
  \BibitemOpen
  \bibfield  {author} {\bibinfo {author} {\bibfnamefont {S.}~\bibnamefont
  {Carr}}, \bibinfo {author} {\bibfnamefont {S.}~\bibnamefont {Fang}}, \bibinfo
  {author} {\bibfnamefont {Z.}~\bibnamefont {Zhu}}, \ and\ \bibinfo {author}
  {\bibfnamefont {E.}~\bibnamefont {Kaxiras}},\ }\href@noop {} {\bibfield
  {journal} {\bibinfo  {journal} {Physical Review Research}\ }\textbf {\bibinfo
  {volume} {1}},\ \bibinfo {pages} {013001} (\bibinfo {year}
  {2019})}\BibitemShut {NoStop}%
\bibitem [{\citenamefont {Shapiro}(1963)}]{shapiro1963josephson}%
  \BibitemOpen
  \bibfield  {author} {\bibinfo {author} {\bibfnamefont {S.}~\bibnamefont
  {Shapiro}},\ }\href@noop {} {\bibfield  {journal} {\bibinfo  {journal}
  {Physical Review Letters}\ }\textbf {\bibinfo {volume} {11}},\ \bibinfo
  {pages} {80} (\bibinfo {year} {1963})}\BibitemShut {NoStop}%
\bibitem [{\citenamefont {Grimes}\ and\ \citenamefont
  {Shapiro}(1968)}]{grimes1968millimeter}%
  \BibitemOpen
  \bibfield  {author} {\bibinfo {author} {\bibfnamefont {C.}~\bibnamefont
  {Grimes}}\ and\ \bibinfo {author} {\bibfnamefont {S.}~\bibnamefont
  {Shapiro}},\ }\href@noop {} {\bibfield  {journal} {\bibinfo  {journal}
  {Physical Review}\ }\textbf {\bibinfo {volume} {169}},\ \bibinfo {pages}
  {397} (\bibinfo {year} {1968})}\BibitemShut {NoStop}%
\bibitem [{\citenamefont {Mudi}\ and\ \citenamefont
  {Frolov}(2021)}]{mudi2021model}%
  \BibitemOpen
  \bibfield  {author} {\bibinfo {author} {\bibfnamefont {S.}~\bibnamefont
  {Mudi}}\ and\ \bibinfo {author} {\bibfnamefont {S.}~\bibnamefont {Frolov}},\
  }\href@noop {} {\bibfield  {journal} {\bibinfo  {journal} {arXiv preprint
  arXiv:2106.00495}\ } (\bibinfo {year} {2021})}\BibitemShut {NoStop}%
\bibitem [{\citenamefont {Stoutimore}\ \emph {et~al.}(2018)\citenamefont
  {Stoutimore}, \citenamefont {Rossolenko}, \citenamefont {Bolginov},
  \citenamefont {Oboznov}, \citenamefont {Rusanov}, \citenamefont {Baranov},
  \citenamefont {Pugach}, \citenamefont {Frolov}, \citenamefont {Ryazanov},\
  and\ \citenamefont {Van~Harlingen}}]{stoutimore2018second}%
  \BibitemOpen
  \bibfield  {author} {\bibinfo {author} {\bibfnamefont {M.}~\bibnamefont
  {Stoutimore}}, \bibinfo {author} {\bibfnamefont {A.}~\bibnamefont
  {Rossolenko}}, \bibinfo {author} {\bibfnamefont {V.}~\bibnamefont
  {Bolginov}}, \bibinfo {author} {\bibfnamefont {V.}~\bibnamefont {Oboznov}},
  \bibinfo {author} {\bibfnamefont {A.}~\bibnamefont {Rusanov}}, \bibinfo
  {author} {\bibfnamefont {D.}~\bibnamefont {Baranov}}, \bibinfo {author}
  {\bibfnamefont {N.}~\bibnamefont {Pugach}}, \bibinfo {author} {\bibfnamefont
  {S.}~\bibnamefont {Frolov}}, \bibinfo {author} {\bibfnamefont
  {V.}~\bibnamefont {Ryazanov}}, \ and\ \bibinfo {author} {\bibfnamefont
  {D.}~\bibnamefont {Van~Harlingen}},\ }\href@noop {} {\bibfield  {journal}
  {\bibinfo  {journal} {Physical review letters}\ }\textbf {\bibinfo {volume}
  {121}},\ \bibinfo {pages} {177702} (\bibinfo {year} {2018})}\BibitemShut
  {NoStop}%
\bibitem [{\citenamefont {Fominov}\ and\ \citenamefont
  {Mikhailov}(2022)}]{fominov2022asymmetric}%
  \BibitemOpen
  \bibfield  {author} {\bibinfo {author} {\bibfnamefont {Y.~V.}\ \bibnamefont
  {Fominov}}\ and\ \bibinfo {author} {\bibfnamefont {D.}~\bibnamefont
  {Mikhailov}},\ }\href@noop {} {\bibfield  {journal} {\bibinfo  {journal}
  {Physical Review B}\ }\textbf {\bibinfo {volume} {106}},\ \bibinfo {pages}
  {134514} (\bibinfo {year} {2022})}\BibitemShut {NoStop}%
\bibitem [{\citenamefont {Souto}\ \emph {et~al.}(2022)\citenamefont {Souto},
  \citenamefont {Leijnse},\ and\ \citenamefont {Schrade}}]{souto2022josephson}%
  \BibitemOpen
  \bibfield  {author} {\bibinfo {author} {\bibfnamefont {R.~S.}\ \bibnamefont
  {Souto}}, \bibinfo {author} {\bibfnamefont {M.}~\bibnamefont {Leijnse}}, \
  and\ \bibinfo {author} {\bibfnamefont {C.}~\bibnamefont {Schrade}},\
  }\href@noop {} {\bibfield  {journal} {\bibinfo  {journal} {Physical Review
  Letters}\ }\textbf {\bibinfo {volume} {129}},\ \bibinfo {pages} {267702}
  (\bibinfo {year} {2022})}\BibitemShut {NoStop}%
\bibitem [{\citenamefont {Lin}\ \emph {et~al.}(2022)\citenamefont {Lin},
  \citenamefont {Siriviboon}, \citenamefont {Scammell}, \citenamefont {Liu},
  \citenamefont {Rhodes}, \citenamefont {Watanabe}, \citenamefont {Taniguchi},
  \citenamefont {Hone}, \citenamefont {Scheurer},\ and\ \citenamefont
  {Li}}]{lin2022zero}%
  \BibitemOpen
  \bibfield  {author} {\bibinfo {author} {\bibfnamefont {J.-X.}\ \bibnamefont
  {Lin}}, \bibinfo {author} {\bibfnamefont {P.}~\bibnamefont {Siriviboon}},
  \bibinfo {author} {\bibfnamefont {H.~D.}\ \bibnamefont {Scammell}}, \bibinfo
  {author} {\bibfnamefont {S.}~\bibnamefont {Liu}}, \bibinfo {author}
  {\bibfnamefont {D.}~\bibnamefont {Rhodes}}, \bibinfo {author} {\bibfnamefont
  {K.}~\bibnamefont {Watanabe}}, \bibinfo {author} {\bibfnamefont
  {T.}~\bibnamefont {Taniguchi}}, \bibinfo {author} {\bibfnamefont
  {J.}~\bibnamefont {Hone}}, \bibinfo {author} {\bibfnamefont {M.~S.}\
  \bibnamefont {Scheurer}}, \ and\ \bibinfo {author} {\bibfnamefont
  {J.}~\bibnamefont {Li}},\ }\href@noop {} {\bibfield  {journal} {\bibinfo
  {journal} {Nature Physics}\ }\textbf {\bibinfo {volume} {18}},\ \bibinfo
  {pages} {1221} (\bibinfo {year} {2022})}\BibitemShut {NoStop}%
\bibitem [{\citenamefont {Scammell}\ \emph {et~al.}(2022)\citenamefont
  {Scammell}, \citenamefont {Li},\ and\ \citenamefont
  {Scheurer}}]{scammell2022theory}%
  \BibitemOpen
  \bibfield  {author} {\bibinfo {author} {\bibfnamefont {H.~D.}\ \bibnamefont
  {Scammell}}, \bibinfo {author} {\bibfnamefont {J.}~\bibnamefont {Li}}, \ and\
  \bibinfo {author} {\bibfnamefont {M.~S.}\ \bibnamefont {Scheurer}},\
  }\href@noop {} {\bibfield  {journal} {\bibinfo  {journal} {2D Materials}\
  }\textbf {\bibinfo {volume} {9}},\ \bibinfo {pages} {025027} (\bibinfo {year}
  {2022})}\BibitemShut {NoStop}%
\bibitem [{\citenamefont {Daido}\ \emph {et~al.}(2022)\citenamefont {Daido},
  \citenamefont {Ikeda},\ and\ \citenamefont {Yanase}}]{daido2022intrinsic}%
  \BibitemOpen
  \bibfield  {author} {\bibinfo {author} {\bibfnamefont {A.}~\bibnamefont
  {Daido}}, \bibinfo {author} {\bibfnamefont {Y.}~\bibnamefont {Ikeda}}, \ and\
  \bibinfo {author} {\bibfnamefont {Y.}~\bibnamefont {Yanase}},\ }\href@noop {}
  {\bibfield  {journal} {\bibinfo  {journal} {Physical Review Letters}\
  }\textbf {\bibinfo {volume} {128}},\ \bibinfo {pages} {037001} (\bibinfo
  {year} {2022})}\BibitemShut {NoStop}%
\bibitem [{\citenamefont {Yuan}\ and\ \citenamefont
  {Fu}(2022)}]{yuan2022supercurrent}%
  \BibitemOpen
  \bibfield  {author} {\bibinfo {author} {\bibfnamefont {N.~F.}\ \bibnamefont
  {Yuan}}\ and\ \bibinfo {author} {\bibfnamefont {L.}~\bibnamefont {Fu}},\
  }\href@noop {} {\bibfield  {journal} {\bibinfo  {journal} {Proceedings of the
  National Academy of Sciences}\ }\textbf {\bibinfo {volume} {119}},\ \bibinfo
  {pages} {e2119548119} (\bibinfo {year} {2022})}\BibitemShut {NoStop}%
\bibitem [{\citenamefont {He}\ \emph {et~al.}(2022)\citenamefont {He},
  \citenamefont {Tanaka},\ and\ \citenamefont
  {Nagaosa}}]{he2022phenomenological}%
  \BibitemOpen
  \bibfield  {author} {\bibinfo {author} {\bibfnamefont {J.~J.}\ \bibnamefont
  {He}}, \bibinfo {author} {\bibfnamefont {Y.}~\bibnamefont {Tanaka}}, \ and\
  \bibinfo {author} {\bibfnamefont {N.}~\bibnamefont {Nagaosa}},\ }\href@noop
  {} {\bibfield  {journal} {\bibinfo  {journal} {New Journal of Physics}\
  }\textbf {\bibinfo {volume} {24}},\ \bibinfo {pages} {053014} (\bibinfo
  {year} {2022})}\BibitemShut {NoStop}%
\bibitem [{\citenamefont {Po}\ \emph {et~al.}(2019)\citenamefont {Po},
  \citenamefont {Zou}, \citenamefont {Senthil},\ and\ \citenamefont
  {Vishwanath}}]{po2019faithful}%
  \BibitemOpen
  \bibfield  {author} {\bibinfo {author} {\bibfnamefont {H.~C.}\ \bibnamefont
  {Po}}, \bibinfo {author} {\bibfnamefont {L.}~\bibnamefont {Zou}}, \bibinfo
  {author} {\bibfnamefont {T.}~\bibnamefont {Senthil}}, \ and\ \bibinfo
  {author} {\bibfnamefont {A.}~\bibnamefont {Vishwanath}},\ }\href@noop {}
  {\bibfield  {journal} {\bibinfo  {journal} {Physical Review B}\ }\textbf
  {\bibinfo {volume} {99}},\ \bibinfo {pages} {195455} (\bibinfo {year}
  {2019})}\BibitemShut {NoStop}%
\bibitem [{\citenamefont {Zhou}\ \emph
  {et~al.}(2021{\natexlab{a}})\citenamefont {Zhou}, \citenamefont {Xie},
  \citenamefont {Taniguchi}, \citenamefont {Watanabe},\ and\ \citenamefont
  {Young}}]{zhou2021superconductivity}%
  \BibitemOpen
  \bibfield  {author} {\bibinfo {author} {\bibfnamefont {H.}~\bibnamefont
  {Zhou}}, \bibinfo {author} {\bibfnamefont {T.}~\bibnamefont {Xie}}, \bibinfo
  {author} {\bibfnamefont {T.}~\bibnamefont {Taniguchi}}, \bibinfo {author}
  {\bibfnamefont {K.}~\bibnamefont {Watanabe}}, \ and\ \bibinfo {author}
  {\bibfnamefont {A.~F.}\ \bibnamefont {Young}},\ }\href@noop {} {\bibfield
  {journal} {\bibinfo  {journal} {Nature}\ }\textbf {\bibinfo {volume} {598}},\
  \bibinfo {pages} {434} (\bibinfo {year} {2021}{\natexlab{a}})}\BibitemShut
  {NoStop}%
\bibitem [{\citenamefont {Zhou}\ \emph
  {et~al.}(2021{\natexlab{b}})\citenamefont {Zhou}, \citenamefont {Xie},
  \citenamefont {Ghazaryan}, \citenamefont {Holder}, \citenamefont {Ehrets},
  \citenamefont {Spanton}, \citenamefont {Taniguchi}, \citenamefont {Watanabe},
  \citenamefont {Berg}, \citenamefont {Serbyn} \emph {et~al.}}]{zhou2021half}%
  \BibitemOpen
  \bibfield  {author} {\bibinfo {author} {\bibfnamefont {H.}~\bibnamefont
  {Zhou}}, \bibinfo {author} {\bibfnamefont {T.}~\bibnamefont {Xie}}, \bibinfo
  {author} {\bibfnamefont {A.}~\bibnamefont {Ghazaryan}}, \bibinfo {author}
  {\bibfnamefont {T.}~\bibnamefont {Holder}}, \bibinfo {author} {\bibfnamefont
  {J.~R.}\ \bibnamefont {Ehrets}}, \bibinfo {author} {\bibfnamefont {E.~M.}\
  \bibnamefont {Spanton}}, \bibinfo {author} {\bibfnamefont {T.}~\bibnamefont
  {Taniguchi}}, \bibinfo {author} {\bibfnamefont {K.}~\bibnamefont {Watanabe}},
  \bibinfo {author} {\bibfnamefont {E.}~\bibnamefont {Berg}}, \bibinfo {author}
  {\bibfnamefont {M.}~\bibnamefont {Serbyn}},  \emph {et~al.},\ }\href@noop {}
  {\bibfield  {journal} {\bibinfo  {journal} {Nature}\ }\textbf {\bibinfo
  {volume} {598}},\ \bibinfo {pages} {429} (\bibinfo {year}
  {2021}{\natexlab{b}})}\BibitemShut {NoStop}%
\bibitem [{\citenamefont {Zhou}\ \emph {et~al.}(2022)\citenamefont {Zhou},
  \citenamefont {Holleis}, \citenamefont {Saito}, \citenamefont {Cohen},
  \citenamefont {Huynh}, \citenamefont {Patterson}, \citenamefont {Yang},
  \citenamefont {Taniguchi}, \citenamefont {Watanabe},\ and\ \citenamefont
  {Young}}]{zhou2022isospin}%
  \BibitemOpen
  \bibfield  {author} {\bibinfo {author} {\bibfnamefont {H.}~\bibnamefont
  {Zhou}}, \bibinfo {author} {\bibfnamefont {L.}~\bibnamefont {Holleis}},
  \bibinfo {author} {\bibfnamefont {Y.}~\bibnamefont {Saito}}, \bibinfo
  {author} {\bibfnamefont {L.}~\bibnamefont {Cohen}}, \bibinfo {author}
  {\bibfnamefont {W.}~\bibnamefont {Huynh}}, \bibinfo {author} {\bibfnamefont
  {C.~L.}\ \bibnamefont {Patterson}}, \bibinfo {author} {\bibfnamefont
  {F.}~\bibnamefont {Yang}}, \bibinfo {author} {\bibfnamefont {T.}~\bibnamefont
  {Taniguchi}}, \bibinfo {author} {\bibfnamefont {K.}~\bibnamefont {Watanabe}},
  \ and\ \bibinfo {author} {\bibfnamefont {A.~F.}\ \bibnamefont {Young}},\
  }\href@noop {} {\bibfield  {journal} {\bibinfo  {journal} {Science}\ }\textbf
  {\bibinfo {volume} {375}},\ \bibinfo {pages} {774} (\bibinfo {year}
  {2022})}\BibitemShut {NoStop}%
\bibitem [{\citenamefont {de~la Barrera}\ \emph {et~al.}(2022)\citenamefont
  {de~la Barrera}, \citenamefont {Aronson}, \citenamefont {Zheng},
  \citenamefont {Watanabe}, \citenamefont {Taniguchi}, \citenamefont {Ma},
  \citenamefont {Jarillo-Herrero},\ and\ \citenamefont
  {Ashoori}}]{de2022cascade}%
  \BibitemOpen
  \bibfield  {author} {\bibinfo {author} {\bibfnamefont {S.~C.}\ \bibnamefont
  {de~la Barrera}}, \bibinfo {author} {\bibfnamefont {S.}~\bibnamefont
  {Aronson}}, \bibinfo {author} {\bibfnamefont {Z.}~\bibnamefont {Zheng}},
  \bibinfo {author} {\bibfnamefont {K.}~\bibnamefont {Watanabe}}, \bibinfo
  {author} {\bibfnamefont {T.}~\bibnamefont {Taniguchi}}, \bibinfo {author}
  {\bibfnamefont {Q.}~\bibnamefont {Ma}}, \bibinfo {author} {\bibfnamefont
  {P.}~\bibnamefont {Jarillo-Herrero}}, \ and\ \bibinfo {author} {\bibfnamefont
  {R.}~\bibnamefont {Ashoori}},\ }\href@noop {} {\bibfield  {journal} {\bibinfo
   {journal} {Nature Physics}\ ,\ \bibinfo {pages} {1}} (\bibinfo {year}
  {2022})}\BibitemShut {NoStop}%
\bibitem [{\citenamefont {Zhang}\ \emph
  {et~al.}(2022{\natexlab{b}})\citenamefont {Zhang}, \citenamefont {Polski},
  \citenamefont {Thomson}, \citenamefont {Lantagne-Hurtubise}, \citenamefont
  {Lewandowski}, \citenamefont {Zhou}, \citenamefont {Watanabe}, \citenamefont
  {Taniguchi}, \citenamefont {Alicea},\ and\ \citenamefont
  {Nadj-Perge}}]{zhang2022spin}%
  \BibitemOpen
  \bibfield  {author} {\bibinfo {author} {\bibfnamefont {Y.}~\bibnamefont
  {Zhang}}, \bibinfo {author} {\bibfnamefont {R.}~\bibnamefont {Polski}},
  \bibinfo {author} {\bibfnamefont {A.}~\bibnamefont {Thomson}}, \bibinfo
  {author} {\bibfnamefont {{\'E}.}~\bibnamefont {Lantagne-Hurtubise}}, \bibinfo
  {author} {\bibfnamefont {C.}~\bibnamefont {Lewandowski}}, \bibinfo {author}
  {\bibfnamefont {H.}~\bibnamefont {Zhou}}, \bibinfo {author} {\bibfnamefont
  {K.}~\bibnamefont {Watanabe}}, \bibinfo {author} {\bibfnamefont
  {T.}~\bibnamefont {Taniguchi}}, \bibinfo {author} {\bibfnamefont
  {J.}~\bibnamefont {Alicea}}, \ and\ \bibinfo {author} {\bibfnamefont
  {S.}~\bibnamefont {Nadj-Perge}},\ }\href@noop {} {\bibfield  {journal}
  {\bibinfo  {journal} {arXiv preprint arXiv:2205.05087}\ } (\bibinfo {year}
  {2022}{\natexlab{b}})}\BibitemShut {NoStop}%
\end{thebibliography}

\begin{thebibliography}{11}%
\makeatletter
\providecommand \@ifxundefined [1]{%
 \@ifx{#1\undefined}
}%
\providecommand \@ifnum [1]{%
 \ifnum #1\expandafter \@firstoftwo
 \else \expandafter \@secondoftwo
 \fi
}%
\providecommand \@ifx [1]{%
 \ifx #1\expandafter \@firstoftwo
 \else \expandafter \@secondoftwo
 \fi
}%
\providecommand \natexlab [1]{#1}%
\providecommand \enquote  [1]{``#1''}%
\providecommand \bibnamefont  [1]{#1}%
\providecommand \bibfnamefont [1]{#1}%
\providecommand \citenamefont [1]{#1}%
\providecommand \href@noop [0]{\@secondoftwo}%
\providecommand \href [0]{\begingroup \@sanitize@url \@href}%
\providecommand \@href[1]{\@@startlink{#1}\@@href}%
\providecommand \@@href[1]{\endgroup#1\@@endlink}%
\providecommand \@sanitize@url [0]{\catcode `\\12\catcode `\$12\catcode
  `\&12\catcode `\#12\catcode `\^12\catcode `\_12\catcode `\%12\relax}%
\providecommand \@@startlink[1]{}%
\providecommand \@@endlink[0]{}%
\providecommand \url  [0]{\begingroup\@sanitize@url \@url }%
\providecommand \@url [1]{\endgroup\@href {#1}{\urlprefix }}%
\providecommand \urlprefix  [0]{URL }%
\providecommand \Eprint [0]{\href }%
\providecommand \doibase [0]{https://doi.org/}%
\providecommand \selectlanguage [0]{\@gobble}%
\providecommand \bibinfo  [0]{\@secondoftwo}%
\providecommand \bibfield  [0]{\@secondoftwo}%
\providecommand \translation [1]{[#1]}%
\providecommand \BibitemOpen [0]{}%
\providecommand \bibitemStop [0]{}%
\providecommand \bibitemNoStop [0]{.\EOS\space}%
\providecommand \EOS [0]{\spacefactor3000\relax}%
\providecommand \BibitemShut  [1]{\csname bibitem#1\endcsname}%
\let\auto@bib@innerbib\@empty
\bibitem [{\citenamefont {Diez-Merida}\ \emph {et~al.}(2021)\citenamefont
  {Diez-Merida}, \citenamefont {D{\'\i}ez-Carl{\'o}n}, \citenamefont {Yang},
  \citenamefont {Xie}, \citenamefont {Gao}, \citenamefont {Watanabe},
  \citenamefont {Taniguchi}, \citenamefont {Lu}, \citenamefont {Law},\ and\
  \citenamefont {Efetov}}]{diez2021magnetic}%
  \BibitemOpen
  \bibfield  {author} {\bibinfo {author} {\bibfnamefont {J.}~\bibnamefont
  {Diez-Merida}}, \bibinfo {author} {\bibfnamefont {A.}~\bibnamefont
  {D{\'\i}ez-Carl{\'o}n}}, \bibinfo {author} {\bibfnamefont {S.}~\bibnamefont
  {Yang}}, \bibinfo {author} {\bibfnamefont {Y.-M.}\ \bibnamefont {Xie}},
  \bibinfo {author} {\bibfnamefont {X.-J.}\ \bibnamefont {Gao}}, \bibinfo
  {author} {\bibfnamefont {K.}~\bibnamefont {Watanabe}}, \bibinfo {author}
  {\bibfnamefont {T.}~\bibnamefont {Taniguchi}}, \bibinfo {author}
  {\bibfnamefont {X.}~\bibnamefont {Lu}}, \bibinfo {author} {\bibfnamefont
  {K.~T.}\ \bibnamefont {Law}},\ and\ \bibinfo {author} {\bibfnamefont {D.~K.}\
  \bibnamefont {Efetov}},\ }\href@noop {} {\bibfield  {journal} {\bibinfo
  {journal} {arXiv preprint arXiv:2110.01067}\ } (\bibinfo {year}
  {2021})}\BibitemShut {NoStop}%
\bibitem [{\citenamefont {Brouwer}\ and\ \citenamefont
  {Beenakker}(1997)}]{brouwer1997anomalous}%
  \BibitemOpen
  \bibfield  {author} {\bibinfo {author} {\bibfnamefont {P.}~\bibnamefont
  {Brouwer}}\ and\ \bibinfo {author} {\bibfnamefont {C.}~\bibnamefont
  {Beenakker}},\ }\href@noop {} {\bibfield  {journal} {\bibinfo  {journal}
  {Chaos, Solitons \& Fractals}\ }\textbf {\bibinfo {volume} {8}},\ \bibinfo
  {pages} {1249} (\bibinfo {year} {1997})}\BibitemShut {NoStop}%
\bibitem [{\citenamefont {Po}\ \emph {et~al.}(2019)\citenamefont {Po},
  \citenamefont {Zou}, \citenamefont {Senthil},\ and\ \citenamefont
  {Vishwanath}}]{po2019faithful}%
  \BibitemOpen
  \bibfield  {author} {\bibinfo {author} {\bibfnamefont {H.~C.}\ \bibnamefont
  {Po}}, \bibinfo {author} {\bibfnamefont {L.}~\bibnamefont {Zou}}, \bibinfo
  {author} {\bibfnamefont {T.}~\bibnamefont {Senthil}},\ and\ \bibinfo {author}
  {\bibfnamefont {A.}~\bibnamefont {Vishwanath}},\ }\href@noop {} {\bibfield
  {journal} {\bibinfo  {journal} {Physical Review B}\ }\textbf {\bibinfo
  {volume} {99}},\ \bibinfo {pages} {195455} (\bibinfo {year}
  {2019})}\BibitemShut {NoStop}%
\bibitem [{\citenamefont {Oh}\ \emph {et~al.}(2021)\citenamefont {Oh},
  \citenamefont {Nuckolls}, \citenamefont {Wong}, \citenamefont {Lee},
  \citenamefont {Liu}, \citenamefont {Watanabe}, \citenamefont {Taniguchi},\
  and\ \citenamefont {Yazdani}}]{oh2021evidence}%
  \BibitemOpen
  \bibfield  {author} {\bibinfo {author} {\bibfnamefont {M.}~\bibnamefont
  {Oh}}, \bibinfo {author} {\bibfnamefont {K.~P.}\ \bibnamefont {Nuckolls}},
  \bibinfo {author} {\bibfnamefont {D.}~\bibnamefont {Wong}}, \bibinfo {author}
  {\bibfnamefont {R.~L.}\ \bibnamefont {Lee}}, \bibinfo {author} {\bibfnamefont
  {X.}~\bibnamefont {Liu}}, \bibinfo {author} {\bibfnamefont {K.}~\bibnamefont
  {Watanabe}}, \bibinfo {author} {\bibfnamefont {T.}~\bibnamefont
  {Taniguchi}},\ and\ \bibinfo {author} {\bibfnamefont {A.}~\bibnamefont
  {Yazdani}},\ }\href@noop {} {\bibfield  {journal} {\bibinfo  {journal}
  {Nature}\ }\textbf {\bibinfo {volume} {600}},\ \bibinfo {pages} {240}
  (\bibinfo {year} {2021})}\BibitemShut {NoStop}%
\bibitem [{\citenamefont {Wu}\ \emph {et~al.}(2018)\citenamefont {Wu},
  \citenamefont {MacDonald},\ and\ \citenamefont {Martin}}]{wu2018theory}%
  \BibitemOpen
  \bibfield  {author} {\bibinfo {author} {\bibfnamefont {F.}~\bibnamefont
  {Wu}}, \bibinfo {author} {\bibfnamefont {A.~H.}\ \bibnamefont {MacDonald}},\
  and\ \bibinfo {author} {\bibfnamefont {I.}~\bibnamefont {Martin}},\
  }\href@noop {} {\bibfield  {journal} {\bibinfo  {journal} {Physical review
  letters}\ }\textbf {\bibinfo {volume} {121}},\ \bibinfo {pages} {257001}
  (\bibinfo {year} {2018})}\BibitemShut {NoStop}%
\bibitem [{\citenamefont {Wu}\ and\ \citenamefont
  {Sarma}(2019)}]{wu2019identification}%
  \BibitemOpen
  \bibfield  {author} {\bibinfo {author} {\bibfnamefont {F.}~\bibnamefont
  {Wu}}\ and\ \bibinfo {author} {\bibfnamefont {S.~D.}\ \bibnamefont {Sarma}},\
  }\href@noop {} {\bibfield  {journal} {\bibinfo  {journal} {Physical Review
  B}\ }\textbf {\bibinfo {volume} {99}},\ \bibinfo {pages} {220507} (\bibinfo
  {year} {2019})}\BibitemShut {NoStop}%
\bibitem [{\citenamefont {Brydon}\ \emph {et~al.}(2019)\citenamefont {Brydon},
  \citenamefont {Abergel}, \citenamefont {Agterberg},\ and\ \citenamefont
  {Yakovenko}}]{brydon2019loop}%
  \BibitemOpen
  \bibfield  {author} {\bibinfo {author} {\bibfnamefont {P.}~\bibnamefont
  {Brydon}}, \bibinfo {author} {\bibfnamefont {D.~S.}\ \bibnamefont {Abergel}},
  \bibinfo {author} {\bibfnamefont {D.}~\bibnamefont {Agterberg}},\ and\
  \bibinfo {author} {\bibfnamefont {V.~M.}\ \bibnamefont {Yakovenko}},\
  }\href@noop {} {\bibfield  {journal} {\bibinfo  {journal} {Physical Review
  X}\ }\textbf {\bibinfo {volume} {9}},\ \bibinfo {pages} {031025} (\bibinfo
  {year} {2019})}\BibitemShut {NoStop}%
\bibitem [{\citenamefont {Zou}\ \emph {et~al.}(2018)\citenamefont {Zou},
  \citenamefont {Po}, \citenamefont {Vishwanath},\ and\ \citenamefont
  {Senthil}}]{zou2018band}%
  \BibitemOpen
  \bibfield  {author} {\bibinfo {author} {\bibfnamefont {L.}~\bibnamefont
  {Zou}}, \bibinfo {author} {\bibfnamefont {H.~C.}\ \bibnamefont {Po}},
  \bibinfo {author} {\bibfnamefont {A.}~\bibnamefont {Vishwanath}},\ and\
  \bibinfo {author} {\bibfnamefont {T.}~\bibnamefont {Senthil}},\ }\href@noop
  {} {\bibfield  {journal} {\bibinfo  {journal} {Physical Review B}\ }\textbf
  {\bibinfo {volume} {98}},\ \bibinfo {pages} {085435} (\bibinfo {year}
  {2018})}\BibitemShut {NoStop}%
\bibitem [{\citenamefont {Das}\ \emph {et~al.}(2021)\citenamefont {Das},
  \citenamefont {Lu}, \citenamefont {Herzog-Arbeitman}, \citenamefont {Song},
  \citenamefont {Watanabe}, \citenamefont {Taniguchi}, \citenamefont
  {Bernevig},\ and\ \citenamefont {Efetov}}]{das2021symmetry}%
  \BibitemOpen
  \bibfield  {author} {\bibinfo {author} {\bibfnamefont {I.}~\bibnamefont
  {Das}}, \bibinfo {author} {\bibfnamefont {X.}~\bibnamefont {Lu}}, \bibinfo
  {author} {\bibfnamefont {J.}~\bibnamefont {Herzog-Arbeitman}}, \bibinfo
  {author} {\bibfnamefont {Z.-D.}\ \bibnamefont {Song}}, \bibinfo {author}
  {\bibfnamefont {K.}~\bibnamefont {Watanabe}}, \bibinfo {author}
  {\bibfnamefont {T.}~\bibnamefont {Taniguchi}}, \bibinfo {author}
  {\bibfnamefont {B.~A.}\ \bibnamefont {Bernevig}},\ and\ \bibinfo {author}
  {\bibfnamefont {D.~K.}\ \bibnamefont {Efetov}},\ }\href@noop {} {\bibfield
  {journal} {\bibinfo  {journal} {Nature Physics}\ }\textbf {\bibinfo {volume}
  {17}},\ \bibinfo {pages} {710} (\bibinfo {year} {2021})}\BibitemShut
  {NoStop}%
\bibitem [{\citenamefont {Lian}\ \emph {et~al.}(2021)\citenamefont {Lian},
  \citenamefont {Song}, \citenamefont {Regnault}, \citenamefont {Efetov},
  \citenamefont {Yazdani},\ and\ \citenamefont {Bernevig}}]{lian2021twisted}%
  \BibitemOpen
  \bibfield  {author} {\bibinfo {author} {\bibfnamefont {B.}~\bibnamefont
  {Lian}}, \bibinfo {author} {\bibfnamefont {Z.-D.}\ \bibnamefont {Song}},
  \bibinfo {author} {\bibfnamefont {N.}~\bibnamefont {Regnault}}, \bibinfo
  {author} {\bibfnamefont {D.~K.}\ \bibnamefont {Efetov}}, \bibinfo {author}
  {\bibfnamefont {A.}~\bibnamefont {Yazdani}},\ and\ \bibinfo {author}
  {\bibfnamefont {B.~A.}\ \bibnamefont {Bernevig}},\ }\href@noop {} {\bibfield
  {journal} {\bibinfo  {journal} {Physical Review B}\ }\textbf {\bibinfo
  {volume} {103}},\ \bibinfo {pages} {205414} (\bibinfo {year}
  {2021})}\BibitemShut {NoStop}%
\bibitem [{\citenamefont {Fu}(2009)}]{fu2009hexagonal}%
  \BibitemOpen
  \bibfield  {author} {\bibinfo {author} {\bibfnamefont {L.}~\bibnamefont
  {Fu}},\ }\href@noop {} {\bibfield  {journal} {\bibinfo  {journal} {Physical
  review letters}\ }\textbf {\bibinfo {volume} {103}},\ \bibinfo {pages}
  {266801} (\bibinfo {year} {2009})}\BibitemShut {NoStop}%
\end{thebibliography}
\end{document}